\documentclass[a4paper,fleqn,usenatbib,useAMS]{mnras}

\usepackage{graphicx}	
\usepackage{amsmath}	
\usepackage{amssymb}	
\usepackage{multicol}        
\usepackage{bm}		
\usepackage{pdflscape}	
\usepackage{threeparttable}

\newcommand{\be}{\begin{equation}}
\newcommand{\ee}{\end{equation}}
\newcommand{\ba}{\begin{eqnarray}}
\newcommand{\ea}{\end{eqnarray}}

\newcommand{\fracb}[2]{\left(\frac{#1}{#2}\right)}

\newcommand\restr[2]{{
		\left.\kern-\nulldelimiterspace 
		#1 
		\vphantom{\big|} 
		\right|_{#2} 
}}

\usepackage{etoolbox}
\makeatletter
\patchcmd\@combinedblfloats{\box\@outputbox}{\unvbox\@outputbox}{}{%
	\errmessage{\noexpand\@combinedblfloats could not be patched}%
}%
\makeatother

\usepackage[T1]{fontenc}
\usepackage{ae,aecompl}

\usepackage{newtxtext,newtxmath}
\usepackage[dvipsnames]{xcolor}

\definecolor{blazeorange}{rgb}{1.0, 0.4, 0.0}
\definecolor{seagreen}{rgb}{0.18, 0.55, 0.34}
\definecolor{rufous}{rgb}{0.66, 0.11, 0.03}
\definecolor{royalfuchsia}{rgb}{0.79, 0.17, 0.57}
\definecolor{scarlet}{rgb}{1.0, 0.13, 0.0}
\definecolor{royalpurple}{rgb}{0.47, 0.32, 0.66}

\title[Lightcurves from Structured Jets]{Afterglow Lightcurves from Misaligned Structured Jets}

\author[Beniamini, Granot, \& Gill (2020)]{
	Paz Beniamini,$^{1}$
	\thanks{Contact e-mail: \href{mailto:paz.beniamini@gmail.com}{paz.beniamini@gmail.com}} 
	Jonathan Granot,$^{2,3}$
	Ramandeep Gill$^{2,3}$
	\\
	$^{1}$Division of Physics, Mathematics and Astronomy, California Institute of Technology, Pasadena, CA 91125, USA\\
	$^{2}$Department of Natural Sciences, The Open University of Israel, 
	1 University Road, PO Box 808, Raanana 4353701, Israel \\
	$^{3}$Department of Physics, The George Washington University, Washington, DC 20052, USA
}

\date{Last updated; in original form}

\pubyear{2020}

\begin{document}
	\label{firstpage}
	\pagerange{\pageref{firstpage}--\pageref{lastpage}}
	\maketitle

	\begin{abstract}
		GRB\,170817A\,/\,GW\,170817 is the first GRB clearly viewed far from the GRB jet's symmetry axis. Its afterglow was densely monitored over a wide range of frequencies and times. It has been modeled extensively, primarily numerically, and although this endeavour was very fruitful, many of the underlying model parameters remain undetermined. We provide analytic modelling of GRB afterglows observed off-axis, considering jets with a narrow core (of half-opening angle $\theta_{\rm c}$) and power-law wings in energy per unit solid angle ($\epsilon=\epsilon_c\Theta^{-a}$ where $\Theta=[1+(\theta/\theta_{\rm c})^2]^{1/2}$) and initial specific kinetic energy ($\Gamma_0-1=[\Gamma_{\rm c,0}-1]\Theta^{-b}$), as well as briefly discuss Gaussian jets. Our study reveals qualitatively different types of lightcurves that can be viewed in future off-axis GRBs, with either single or double peaks, depending on the jet structure and the viewing angle. Considering the lightcurve shape rather than the absolute normalizations of times and / or fluxes, removes the dependence of the lightcurve on many of the highly degenerate burst parameters. This study can be easily used to determine the underlying jet structure, significantly reduce the effective parameter space for numerical fitting attempts and provide physical insights. As an illustration, we show that for GRB 170817A, there is a strong correlation between the allowed values of $\Gamma_{\rm c,0}$ and $b$, leading to a narrow strip of allowed solutions in the $\Gamma_{\rm c,0}$-$b$ plane above some minimal values $\Gamma_{\rm c,0}\gtrsim 40, b\gtrsim1.2$. Furthermore, the Lorentz factor of the material dominating the early lightcurve can be constrained by three independent techniques to be $\Gamma_{0}(\theta_{\rm min,0})\approx5-7$.
	\end{abstract}
	
	
	
	\section{Introduction}
	The detection of a binary neutron star merger with LIGO, GW170817 \citep{GW170817}, accompanied by a long lived GRB afterglow 
	\citep[e.g.,][]{Abbott+17-GW170817A-MMO} has enabled us for the first time to unambiguously observe the afterglow of a GRB seen 
	from latitudes much greater than the jet's core.
	The late peak and the slow rise of the lightcurve towards that peak have been modelled as arising due to either the angular \citep{lamb2017,Lazzati2017,Granot2017ApJ...850L..24G,Gill-Granot-18,Kathirgamaraju2018} or radial \citep{Kasliwal2017,Gill-Granot-18,Gottlieb2018,NP2018} structure of the outflow. The observation of superluminal motion \citep{Mooley2018,Ghirlanda2019}, as well as the sharp decline of the lightcurve after the peak \citep[e.g.,][]{Margutti2018,DAvanzo2018,Troja2018,LMR2018,Hajela2019,Lamb2019} suggests the flow had an energetic and relativistic compact core rather than a quasi-spherical structure, implying that an angular structured jet is required to explain, at least the late time afterglow observations.
	
	Determining the angular structure of GRB jets outside of their cores is of crucial importance for advancing our knowledge of various phenomena, such as the physics of formation and dynamics of relativistic jets \citep{Kathirgamaraju2018,Granot2018,BBPG2020}, the underlying mechanism powering the prompt phase of GRBs \citep{BPBG2019,BN2019} and possibly some phenomena observed in cosmological GRB afterglows, like X-ray plateaus \citep{Eichler-Granot-06,BDDM2020,Oganesyan2019}.

	Numerical modelling of off-axis GRB jets with an angular structure has been extensively studied in the literature in order to try and determine the underlying physical properties of the burst and the surrounding medium \citep[e.g.,][]{alexander2018,lamb2018,Lazzati2018,Xie2018,Ryan2019}. Although such modelling is extremely useful it encounters two significant limitations. The first is the large parameter space that must be explored in order to determine those properties. This results in significant computational costs involved in thorough modelling attempts. The second limitation is due to the degeneracies between the intrinsic parameters in their effects on the final lightcurves. Both of these limitations can be largely overcome with the aid of analytical modelling. The latter can reduce the effective parameter space and point towards the unknown parameters (or combinations of those) that may be well constrained by observations \citep[e.g.][]{Gill+19}.

	Analytical modelling of GRB afterglows arising from a jet with angular structure have been carried out by various authors \citep[e.g.][]{Rossi02,Kumar-Granot-03,Granot-Kumar-03,PK03,Rossi04,Granot-RamirezRuiz-Perna-05,Eichler-Granot-06}. Here, motivated by GRB 170817A, and the expectation of seeing future similarly off-axis GRBs triggered by GW detections, we systematically consider the afterglow lightcurves for GRBs that are viewed far from the jet axis. Analytical treatment of this situation reveals qualitatively different types of evolution that may be seen by different observers. In particular we find that either single or double peaked lightcurves can be obtained and the distinction between the two is directly related to the jet structure and the observer's viewing angle. 
	
	The paper is organized as follows. In \S \ref{sec:definitions} we introduce the basic definitions of the jet properties including its angular structure, and describe the dynamics underlying the GRB afterglow as viewed in observer frame coordinates. In \S \ref{sec:theta}, we describe the contributions from different regions (or latitudes) in the jet to the observed emission and their temporal evolution. In \S \ref{sec:Modelling} we examine the different types of afterglow lightcurves that can be viewed depending on the jet properties and the observation angle. We present in \S \ref{sec:Inferences} the relation between the observable properties of off-axis GRB afterglows and the underlying physical properties and show which of the latter can be robustly determined by the former. In \S \ref{sec:170817} we demonstrate the effectiveness of this technique on GRB 170817A. We finally conclude in \S \ref{sec:conc}.
	
	\section{Power-Law Structured Jet: Dynamics and Scalings}
	\label{sec:definitions}
	For simplicity we consider a jet angular structure consisting of a narrow core of half-opening angle $\theta_{\rm c}$ that smoothly transitions into power-law wings in both the kinetic energy per unit solid angle, $\epsilon\equiv dE_{\rm k}/d\Omega$, and the initial specific kinetic energy, $\Gamma_0-1$. For concreteness, we will mainly focus in this paper on power-law (PL) profiles for both quantities (following \citealt{Granot-Kumar-03}) but will briefly discuss implications for Gaussian jets in \S \ref{sec:Gauss}. In the former case, the energy and Lorentz factor are given by
	\begin{equation}\label{eq:PLJ}
	\frac{\epsilon(\theta)}{\epsilon_{\rm c}} = \Theta^{-a}~,\quad\frac{\Gamma_0(\theta)-1}{\Gamma_{\rm c,0}-1} = \Theta^{-b}~,
	\quad\quad\Theta \equiv \sqrt{1+\displaystyle\fracb{\theta}{\theta_{\rm c}}^2}\ ,
	\end{equation}
	where $\theta$ is the polar angle measured from the jet's symmetry axis ($\theta_{\rm obs}$ is the polar angle of the observer's line-of-sight). We also define $\xi_{\rm c}\equiv(\Gamma_{\rm c,0}\theta_{\rm c})^2$ and $q\equiv\theta_{\rm{obs}}/\theta_{\rm c}$ such that $\Theta_{\rm{obs}}=\sqrt{1+q^2}$, which will be useful quantities later on. We consider emission form a thin shell behind the external shock that is radially located at $R(\theta,t)$, where each point on the jet expands as if part of a spherical flow with no lateral 
	expansion.
	
	To a first approximation, and as long as the flow is still ultra-relativistic, $\Gamma(\theta)$ may be assumed to be constant up to the deceleration radius $R_{\rm dec}(\theta)$, after which it starts decelerating as a PL in radius,
	\begin{equation}
	\label{eq:GammaR}
	\Gamma(\theta,R)=\Gamma_0(\theta)\times\left\{ \begin{array}{ll}1 & R<R_{\rm dec}(\theta)\,\\
	\zeta^{k-3\over 2} & R>R_{\rm dec}(\theta)\ ,
	\end{array} \right.
	\end{equation}
	where $\zeta\equiv R/R_{\rm dec}(\theta)$, $k$ relates the external density to the radius, i.e.  $\rho= AR^{-k}$ and where the values of the deceleration radii, $R_{\rm dec}(\theta)$, can conveniently be scaled compared to that at the core
	\begin{eqnarray}\label{eq:R_dec}
	R_{\rm{dec},c}&=&\left[\frac{(3-k)\epsilon_c}{Ac^2\Gamma_{\rm c,0}^2}\right]^\frac{1}{3-k}\ ,\quad \frac{R_{\rm{dec}}(\theta)}{R_{\rm{dec},c}}=\Theta^\frac{2b-a}{3-k}\ . 
	\end{eqnarray}

	A slightly more complex expression than Eq.~(\ref{eq:GammaR}), that is valid also for $\Gamma\gtrsim 1$ and self-consistently accounts 
	for energy conservation is given by \citep{PK2000,Gill-Granot-18}
	\begin{equation}
	\label{eq:GammaPK}
	\Gamma(\theta,R)=\frac{\Gamma_0(\theta)+1}{2}\zeta^{k-3}\Bigg[\sqrt{1+\frac{4\Gamma_0(\theta)}{\Gamma_0(\theta)+1}\zeta^{3-k}+\bigg(\frac{2\zeta^{3-k}}{\Gamma_0(\theta)+1}\bigg)^2}-1 \Bigg]
	\end{equation}
	Both Eqs.~(\ref{eq:GammaR}) and (\ref{eq:GammaPK}) hold as long as the dynamics are completely radial, i.e. in the limit of no lateral expansion, since for simplicity we assume that $\epsilon(\theta)$ does not evolve with time and at each $\theta$ the flow behaves as if it were part of a spherical flow with the local value of $\epsilon(\theta)$.
	This approximation is expected to hold so long as\footnote{More accurately, the local condition that allows for significant lateral expansion from causality considerations is $\Gamma\theta<1$. This may occur in some parts of the jet, while others are still ``frozen" to their initial energy. As long as the jet is relativistic then the change in $\epsilon(\theta)$ is typically not very large and can still be neglected \citep{Kumar-Granot-03}.} $\Gamma_{\rm c}>\theta_{\rm c}^{-1}$, as before that point the core of the jet is causally disconnected from the wings. Beyond that radius the jet can begin to expand sideways, causing the radial velocity to decrease and the energy structure to be modified.
	
	The degree to which lateral spreading occurs is still a topic of debate, and different formulations have been proposed in the literature (e.g. \citealt{Rhoads1999,PM1999,SPH1999,Granot-Piran-12}), which apply to a top-hat jet rather than a structured jet. For the sake of clarity and to avoid the uncertainties that are involved in the expectations from lateral spreading we focus here on the situation where the lateral spreading is negligible (in particular, this enables us to directly compare the results to semi-analytic models). The general expectation is that if jet's lateral spreading becomes important, it will mostly affect observers at large viewing angles, for which it would cause the main peak of the lightcurve to occur earlier, with a steeper rise leading to this peak and a steeper decay following it.
	
	By integrating Eq.~(\ref{eq:GammaPK}) we can find the (source frame) emission time $t_{\rm s}(\theta,R) =\int_0^{R} dR'/ (c\beta(\theta,R'))$, where $\beta(\theta,R)\equiv\sqrt{1-\Gamma(\theta,R)^{-2}}$ is the normalized velocity corresponding to $\Gamma(\theta,R)$. The source frame time is related to the observer time via the light travel time from different locations and for points in the jet that are along the line connecting the jet axis and the line  of sight is given by 
	\begin{equation}
	\label{eq:EATS}
	t=t_{\rm s}-R\cos (\theta_{\rm obs}-\theta)/c\,,
	\end{equation}
	where here and in what follows, we omit redshift corrections for clarity (these can be trivially added retrospectively and in any case are expected to be small for off-axis events discussed here).
	For an observer within the beaming cone of the material from each angle $\theta$ ($|\theta_{\rm obs}-\theta|\lesssim1/\Gamma(\theta,t)$), $t(\theta,R)=t_{\rm s}(\theta,R)/2\Gamma(\theta,R)^2$.
	We define the apparent deceleration times for such observers
	\begin{eqnarray}
	\label{eq:t_dec}
	t_{\rm{dec},c}&=&\frac{R_{\rm{dec},c}}{2c\Gamma_{\rm c,0}^2}\ ,\quad \frac{t_{\rm{dec}}(\theta)}{t_{\rm{dec},c}}=\Theta^\frac{2(4-k)b-a}{3-k}\ .
	\end{eqnarray}
	After the deceleration time and while the flow is still relativistic ($t_{\rm dec}(\theta)<t<t_{\rm NR}(\theta)$) we obtain an approximation for a PL jet,
	\begin{equation}
	\label{eq:Gammathetat}
	\Gamma(\theta,t) =\Gamma_{\rm c,0}\fracb{t}{t_{\rm dec,c}}^\frac{k-3}{8-2k}\left[1+\fracb{\theta}{\theta_{\rm c}}^2\right]^\frac{-a}{4(4-k)}
	= \Gamma_{\rm c,0}\tilde{t}^\frac{k-3}{8-2k}\Theta^\frac{-a}{8-2k}\ ,
	\end{equation}
	where $\tilde{t}\equiv t/t_{\rm dec,c}$.

	\begin{table}
		\caption{Some useful notations considered in this paper.}
		\resizebox{0.5\textwidth}{!}{
			\begin{threeparttable}
				\begin{tabular}{ccc}\hline
					Notation &  Definition & Relevant  \\
					& & equation\\ \hline
					$\theta_{\rm c}$ & Half-opening angle of the jet's core &  -- \\
					$\theta_{\rm obs}$ & Observer's viewing angle (relative to jet axis) &  -- \\
					$q$ & Normalized observer's viewing angle, $\theta_{\rm obs}/\theta_{\rm c}$ &   -\\
					$\Theta$ & $\sqrt{1+q^2}$ &   \ref{eq:PLJ}\\
					$\theta_{F}(t)$ & Polar angle of matter dominating $F_{\nu}(t)$ & \ref{eq:dFnudOmega} \\
					$\theta_{\rm min}(t)$ & Lowest latitude within $\Gamma^{-1}(\theta_{\rm min})$ from observer &  \ref{eq:thetaminfull} \\
					$\theta_{*}$ & Lowest latitude {\it initially} beamed to the observer &   \ref{equation:theta*} \\
					$\theta_{\rm beam}(t)$ & Polar angle  equal to its beaming 
					angle, $\Gamma\theta=1$ & \ref{eq:thetabeam}\\
					$\theta_{\rm dec}(t)$ & Latitude decelerating at $t$ &   \ref{eq:thetadec}\\
					\hline 
					$\epsilon_{\rm c}$ & Jet core's initial kinetic energy per solid angle &   \ref{eq:PLJ}\\
					$\Gamma_{\rm c,0}$ & Jet core's initial Lorentz factor &   \ref{eq:PLJ}\\
					$\xi_{\rm c}$ & $(\Gamma_{\rm c,0}\theta_{\rm c})^2$ &   -\\
					$\zeta$ & $R/R_{\rm dec}(\theta)$ &   -\\
					$a$ & 
					Jet's energy angular slope: $-d\log\epsilon/d\log\Theta$ & \ref{eq:PLJ} \\
					$b$ & 
					Initial specific kinetic energy angular slope: & \ref{eq:PLJ} \\
					& $-d\log(\Gamma_0-1)/d\log\Theta$ & \\
					$k$ & External density power-law index: $\rho\!=\!A R^{-k}$ & - \\
					\hline  
					$t_{\rm dec,c}$ & Apparent deceleration time of the jet's core & \ref{eq:t_dec} \\
					$\tilde{t}$ & Normalized (apparent) time, $ t/t_{\rm dec,c}$ & -\\
					$\tilde{t}_*$ & $ \tilde{t}_{\rm dec}(\theta_*)$ & \ref{eq:t*}\\
					$\tilde{t}_{\rm c}$ & $\Gamma_{\rm c}(\tilde{t}_{\rm c})=\theta_{\rm c}^{-1}$ & -\\
					$\tilde{t}_{\rm pk}$ & Normalized time of main peak & \ref{eq:t_pk}\\
					$\tilde{t}_{\rm 1pk}$ & Normalized time of early peak & \ref{eq:t1pk}\\
					$\tilde{t}_{\rm dip}$ & Normalized time of dip & \ref{eq:tdip}\\
					\hline 
					$F_{\rm pk}$ & Flux density of main peak & \ref{eq:Fpk}  \\
					$F_{\rm 1pk}$ & Flux density of early peak & \ref{eq:F1pk}\\
					\label{tbl:definitions}
				\end{tabular}
			\end{threeparttable}
		}
	\end{table}

	\section{Regions dominating the observed emission}
	\label{sec:theta}
	For energy structures that are reasonably steep ($a\gtrsim 2$), more inner regions of the jet, that have lower $\theta$, can potentially result in larger contributions to the emission (provided that their radiation is not beamed away from the observer and that the material there has begun decelerating and therefore radiating significantly). It is therefore constructive to define two characteristic angles:  (i) $\theta_{\rm min}(t,\theta_{\rm obs})$ as the minimal polar angle that becomes visible to an observer at $\theta_{\rm obs}$ (i.e., the observer enters the $\Gamma^{-1}$ beaming cone from $\theta_{\rm min}$) at time $t$ (following \citealt{Gill-Granot-18}), and (ii) $\theta_{F}(t,\theta_{\rm obs})$ 
	as the angle that dominates the contribution to the flux received by an observer at $\theta_{\rm obs}$ at time $t$ \citep{Takahashi2019}.
	A summary of all the characteristic angles and other notations in the problem is provided in Table~\ref{tbl:definitions}.
	
	The angle $\theta_{\rm min}(t,\theta_{\rm obs})$ is given by
	\begin{equation}
	\label{eq:thetaminfull}
	\theta_{\rm obs}-\theta_{\rm min}=\frac{1}{\Gamma(\theta_{\rm min},t)}=\left[\frac{2^{3-k}A c^{5-k}t^{3-k}}{(3-k)\epsilon_{\rm c}}\right]^\frac{1}{8-2k} \Theta_{\rm min}^\frac{a}{8-2k}
	\end{equation}
	where the second equality is valid for a PL jet and $t>t_{\rm dec}(\theta_{\rm min})$.

	To find $\theta_{F}(\theta_{\rm obs},t)$ one needs to maximize the contribution to $dF_{\nu}(\theta_{\rm obs},t)/d\Omega$ as a function of $\theta$. Because of azimuthal symmetry, the corresponding brightest point of the jet is always along the line connecting the jet axis and the line of sight, so we can use Eq.~(\ref{eq:EATS}), and more generally also $\hat{n}\cdot\hat\beta=\hat{n}\cdot\hat{r}=\cos\tilde\theta=\cos(\theta_{\rm obs}-\theta)$, where $\tilde\theta$ is the angle from the line of sight. Recall that $dF_\nu/d\Omega \propto \mathcal{D}^3 \frac{dL'}{d\nu'}$ where $\mathcal{D}(\theta,t)=[\Gamma(1-\beta \cos(\theta_{\rm obs}-\theta))]^{-1}$ is the Doppler factor and $\frac{dL'}{d\nu'}(\theta,t)$ is the spectral luminosity in the jet's co-moving frame. For a PL spectrum in the co-moving frame: $\frac{dL'}{d\nu'}\propto (\nu')^{-\beta_\nu}$, one obtains $dF_\nu/d\Omega \propto \mathcal{D}^{3+\beta_{\nu}} \restr{\frac{dL'}{d\nu'}}{\nu}$.
	Assuming the spectrum to be dominated by synchrotron radiation from the forward shock, we can express $\restr{\frac{dL'}{d\nu'}}{\nu}$ in terms of $\epsilon,R$ and $\beta_{\nu}$ in terms of $p$ (the slope of the accelerated electrons energy power-law distribution, with $dN/d\gamma_e\propto\gamma_e^{-p}$ for $\gamma_e>\gamma_m$). For example for $R>R_{\rm dec}(\theta)$ and $\nu_c>\nu_m$ we obtain\footnote{This is expected to be the case starting from relatively early times. Similar expressions can be found for the three other cases, following the expressions for $\restr{\frac{dL'}{d\nu'}}{\nu}$ as a function of $\zeta$ as detailed in \cite{Granot05} and Table~\ref{tbl:Lambda}.}
	\begin{eqnarray}
	\label{eq:dFnudOmega}
	\frac{dF_{\nu}(\theta,R)}{d\Omega}\propto \left\{ \begin{array}{ll}\mathcal{D}^{8/3}R^{3-4k/3} & \nu_a<\nu<\nu_m\ ,\\
	\mathcal{D}^{5+p\over 2}\epsilon^{3p-1\over 4}R^{[15-9p-2k(3-p)]/4} & \nu_m<\nu<\nu_c\ ,
	\\
	\mathcal{D}^{6+p\over 2}\epsilon^{3p-2\over 4}R^{[14-9p+2k(p-2)]/4} & \nu_m\,,\,\nu_c<\nu\ ,
	\end{array} \right.
	\end{eqnarray}
	where $\nu_a$ is the synchrotron self-absorption frequency, $\nu_m$ is the synchrotron emission frequency of minimal energy ($\gamma_e=\gamma_m$) electrons, and 
	$\nu_c$ is that of electrons that cool on the dynamical time. We obtain an approximation for $\theta_{F}(\theta_{\rm obs},t)$ in the following way. We first find $\theta_{F}(R)$ by maximizing $dF_{\nu}(\theta,R)/d\Omega$ over $\theta$. We then relate $R$ and $\theta_{F}(R)$ to the observer frame using the relation for the equal arrival time given in Eq.~(\ref{eq:EATS}), $t=t_{\rm s}(\theta_{F}(R),R)-R\cos (\theta_{\rm obs}-\theta_{F}(R))/c$
	This is an approximation, as we are maximizing for a constant $R$ rather than a constant $t$ or a full integration over the equal arrival time surface.
	This procedure can be easily and rapidly evaluated numerically.
	The goodness of this approximation can be evaluated by comparing to the numerical model presented in \cite{Gill-Granot-18} hereafter GG18). The latter involves a full integration of the flux over the entire jet surface at all emission times and frequencies.

\begin{figure}
		\includegraphics[width=0.45\textwidth]{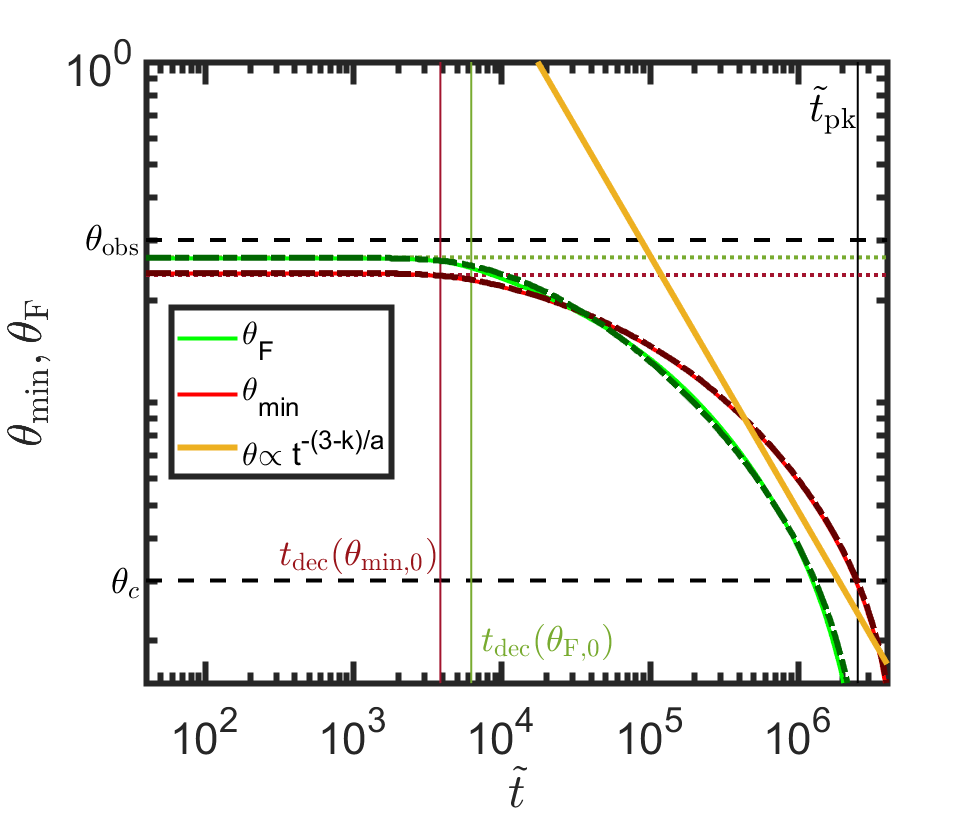}
		\caption{Temporal evolution of the angle from which the observed flux is dominated ($\theta_{F}(t)$ ; green) and the minimal angle at the edge of the beaming cone from the observer ($\theta_{\rm min}(t)$ ; red) as a function of $\tilde{t}\equiv t/t_{\rm dec,c}$. Dot dashed lines depict the numerical approximations based on the formulation presented in \S\;\ref{sec:theta} while solid lines depict the results extracted from the numerical model of GG18. Horizontal dashed lines depict the estimates to $\theta_{F,0},\theta_{\rm min,0}$ given by equations \ref{eq:thetaFprox}, \ref{eq:thetaFdef} respectively. A solid yellow line depicts the approximate PL evolution expected at late times (equation \ref{eq:thetamin}). All cases are plotted for $\Gamma_{\rm c}=1000,\theta_{\rm c}=0.03,a=4,b=2,\theta_{\rm obs}=0.3,k=0,p=2.2$ and assuming power-law segment (PLS) G (see \citealt{GS02})} for the synchrotron emission.
		\label{fig:thetaminthetaF}
	\end{figure}

\begin{figure*}
		\centering
		\includegraphics[width=0.96\textwidth]{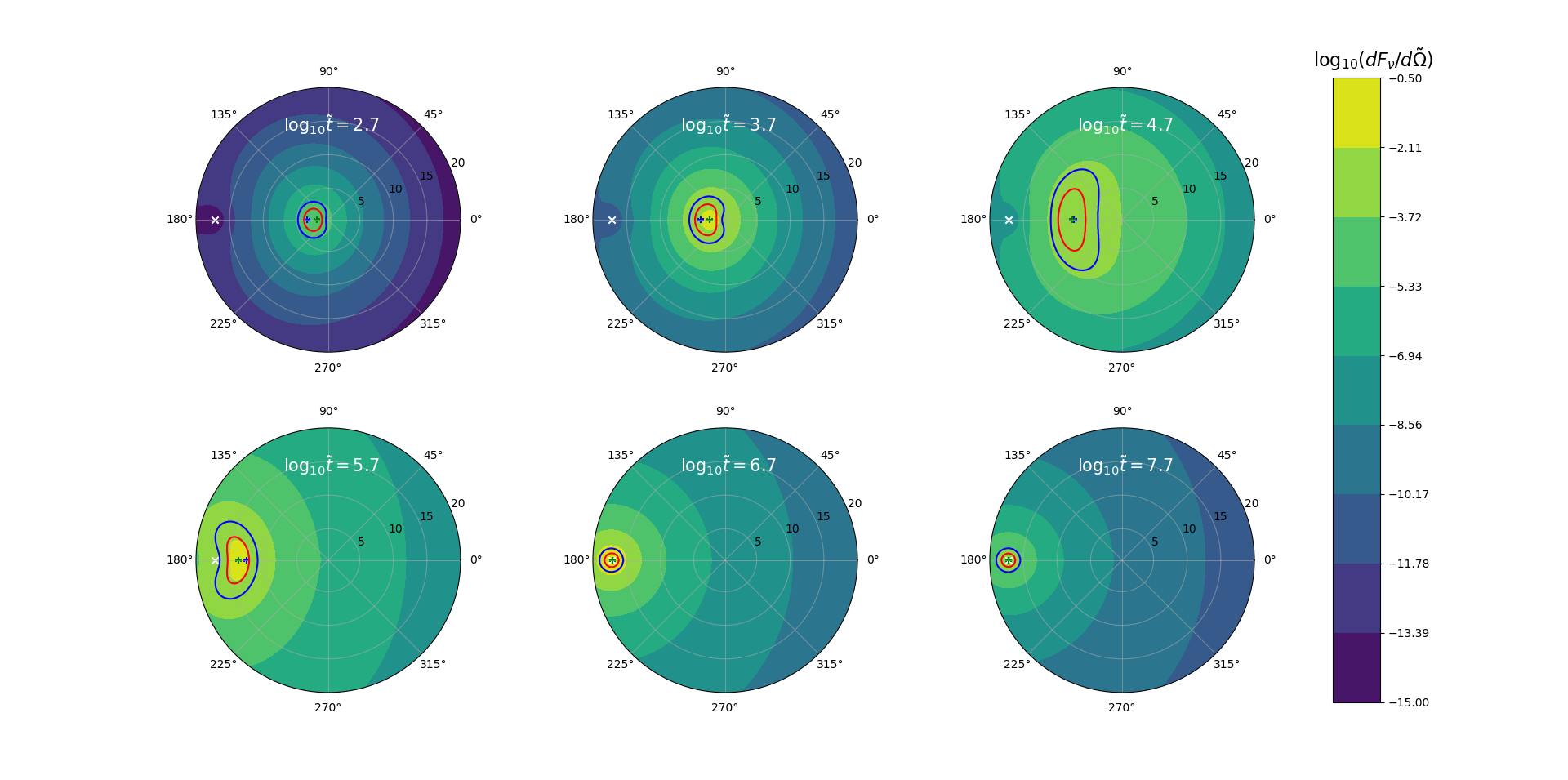}
		\caption{Angular map of $dF_{\nu}/d\tilde{\Omega}$ at different times, where $d\tilde{\Omega}$ denotes the solid angle centered around the line of sight to the observer. A white cross marks the axis of the jet. The peak of $dF_{\nu}/d\tilde{\Omega}(\tilde{t})$ is $\theta_{\rm obs}-\theta_{F}(\tilde{t})$ and is marked with a green `+' sign. Red and blue contours contain angular regions contributing 50$\%$ and 80$\%$ of the total flux, respectively. For comparison we show with a blue `+' sign the value of $\theta_{\rm obs}-\theta_{\rm min}(\tilde{t})$. Results are plotted using the method outlined in GG18 for $\Gamma_{\rm c}=1000,\theta_{\rm c}=0.03,a=4,b=2,\theta_{\rm obs}=0.3,k=0,p=2.2$ and for a frequency such that the emission is dominated by PLS G of the synchrotron emission.}
		\label{fig:dFnudOmegamap}
	\end{figure*}

	A comparison of our approximations for $\theta_{F}(t),\theta_{\rm min}(t)$ to the values extracted from the full integration from the calculation of GG18 is shown in Fig. \ref{fig:thetaminthetaF}. We also present the angular maps showing the strength of $dF_{\nu}/d\tilde\Omega$, where $d\tilde\Omega$ is the unit solid angle 
	measured around the line-of-sight, in Fig. \ref{fig:dFnudOmegamap}. 
	This figure demonstrates that at early and\,/\,or late times the image of the source is roughly spherical around $\theta_{F}(t)$, while at intermediate times, the flux contours tend to deviate from the spherical assumption that we make later. We return to address the importance of this fact in \S\;\ref{sec:Modelling}.

	Both $\theta_{F}(t),\theta_{\rm min}(t)$ are initially roughly constant, with
	$ \theta_{\rm min,0}<\theta_{F,0}<\theta_{\rm obs}$, where $\theta_{\rm min}(t=0)\equiv \theta_{\rm min,0}, \theta_{F}(t=0)\equiv \theta_{F,0}$.
	
	One may be obtain $\theta_{\rm min,0}$ by replacing $\Gamma$ with $\Gamma_0$ in Eq.~(\ref{eq:thetaminfull}) which for\footnote{This condition is required in order for relativistic beaming to be important early on, and for the approximation of the beaming-cone half-opening angle as $1/\Gamma$ to hold. Note that depending on the jet structure, the limit $\Gamma_0(\theta_{\rm min,0})\gg1$ may not hold for large viewing angles.
	} 
	$\Gamma_0(\theta_{\rm min,0})\gg1$ yields 
	\begin{equation}
	\label{eq:x}
	\xi_{\rm c}(q-x)^2=(1+x^2)^b\quad,\quad x\equiv\theta_{\rm min,0}/\theta_{\rm c}\ .
	\end{equation}
	For $b=1,2$ this has the relatively simple analytic solutions,
	\begin{eqnarray}
	\label{eq:thetaFdef}
	&\frac{\theta_{\rm min,0}}{\theta_{\rm c}} = \frac{\xi_{\rm c} q - \sqrt{\xi_{\rm c} q^2 +\xi_{\rm c}-1}}{\xi_{\rm c}-1}\ \mbox{ for } b=1
	\\ & \frac{\theta_{\rm min,0}}{\theta_{\rm c}} = \frac{1}{2}\left(\sqrt{4\sqrt{\xi_{\rm c}}q+\xi_{\rm c}-4}-\sqrt{\xi_{\rm c}}\right)\ \mbox{ for } b=2
	\end{eqnarray}
	
	For $\theta_{F,0}$, the equation in the ultra-relativistic limit is given by
	\begin{equation}
	\label{eq:thetaFprox}
	(2b-\Lambda)(q\!-\!y)^2+2\frac{1+y^2}{y}(q\!-\!y)=\frac{\Lambda}{\xi_{\rm c}}(1+y^2)^{b} \quad,\quad y\equiv\frac{\theta_{F,0}}{\theta_{\rm c}}
	\end{equation}
	where 
	\begin{equation}
	\label{eq:Lambda}
	\Lambda=b+a\frac{ \lambda_{\epsilon}}{\lambda_{\mathcal{D}}}
	\end{equation}
	and $\lambda_{\epsilon},\lambda_{\mathcal{D}}$ are the power-law exponents of $\epsilon,\mathcal{D}$ respectively in Eq.~(\ref{eq:dFnudOmega}). 
	For example, for $\zeta>1,\nu_m<\nu<\nu_c$ (PLS G of \citealt{GS02}), we get $\Lambda=b+\frac{a (3p-1)}{2(5+p)}$. A list of values relevant for other regimes is given in Table~\ref{tbl:Lambda}.
	Approximate solutions to Eq. \ref{eq:thetaFprox} can be given in two limiting cases, depending on the value of $\theta_{\rm obs}$ relative to the critical angle $\theta_*=\theta_{\rm c} \xi_{\rm c}^{1\over 2(b-1)}$ (see Eq. \ref{equation:theta*} and \S\ref{sec:Modelling}) for a physical interpretation of $\theta_*$). The result is $\theta_{F,0}\approx \theta_{\rm obs}$ for $\theta_{\rm obs}\ll \theta_*$ and $\theta_{F,0}\approx \theta_*(\theta_{\rm obs}/\theta_*)^{1/b}$ for $\theta_{\rm obs}\gg \theta_*$ (these two limits can be understood intuitively, see \S \ref{sec:case1}).
	One can combine these limits into an approximation that can be used without prior knowledge of $\theta_{F,0}$:
	\begin{equation}
	\label{eq:thetaF0full}
	\theta_{F,0}=\bigg[\theta_{\rm obs}^{-s}+\bigg(\theta_*\bigg(\frac{\theta_{\rm obs}}{\theta_{*}}\bigg)^{1\over b}\bigg)^{-s}\bigg]^{-1/s}
	\end{equation}
	where $s>0$ is a smoothing parameter that ensures the transition between the appropriate approximations at $\theta_{\rm obs} \approx \theta_*$. A choice that matches well the exact solution is $s=1.5$.
	
	\begin{table*}
		\centering
		\caption{Value of $\Lambda$ as defined by Eq.~(\ref{eq:Lambda}) for different ranges of $\zeta\!=\!R/R_{\rm dec}(\theta)$ and observed synchrotron PLSs, using the notations introduced by \citealt{GS02} in brackets. We also quote the spectral index in each PLS ($\beta_{\nu}$) and the temporal index $\alpha_i$ for a spherical outflow (
			$\alpha_r$ for $\zeta<1$ or $\alpha_d$ for $\zeta\!>\!1$).}
		\centering
		\resizebox{0.7\textwidth}{!}{
			\begin{threeparttable}	\begin{tabular}{ccccccc}\hline	
					$\Lambda$ &  $\zeta$ & PLS & $\beta_{\nu}$ & $\alpha_i$  & $\lambda_{\mathcal{D}}$ & $\lambda_{\epsilon}$ \\  \hline\vspace{0.1cm} 
					$b$ &$\zeta>1$ &  $\nu_a<\nu<\nu_m<\nu_c$ (D) & -1/3  & $\frac{2-k}{4-k}$ & $8/3$ & $0$ \\ \vspace{0.1cm}
					$b+\frac{a}{4}$ &$\zeta>1$ &  $\nu_a<\nu<\nu_c<\nu_m$ (E) & -1/3 & $\frac{2-3k}{3(4-k)}$ & $8/3$ & $2/3$ \\ \vspace{0.1cm} 
					$b+\frac{a}{14}$ &$\zeta>1$ &  $\nu_c<\nu<\nu_m$ (F) & 1/2 & $-1/4$ & $7/2$ & $1/4$ \\ \vspace{0.1cm} 
					$b+\frac{a(3p-1)}{2(5+p)}$ &$\zeta>1$ &  $\nu_m<\nu<\nu_c$ (G) & (p-1)/2 & $\frac{k(3p-5)-12(p-1)}{4(4-k)}$ & $\frac{p+5}{2}$ & $\frac{3p-1}{4}$ \\
					$b+\frac{a(3p-2)}{2(6+p)}$ &$\zeta>1$ &  $\nu>\max(\nu_m,\nu_c)$ (H) & p/2 & $(2-3p)/4$ & $\frac{p+6}{2}$ & $\frac{3p-2}{4}$ \\
					\hline  
					$b$ &$\zeta<1$ &  $\nu_a<\nu<\nu_m<\nu_c$ (D) & -1/3  & $3-k/2$ & $8/3$ & $0$ \\
					$b$ &$\zeta<1$ &  $\nu_a<\nu<\nu_c<\nu_m$ (E) & -1/3 & $11/3-2k$ & $8/3$ & $0$ \\
					$b$ &$\zeta<1$ &  $\nu_c<\nu<\nu_m$ (F) & 1/2 & $2-3k/4$ & $7/2$ & $0$ \\
					$b$ &$\zeta<1$ &  $\nu_m<\nu<\nu_c$ (G) & (p-1)/2 & $3-k(p+5)/4$ & $\frac{p+5}{2}$ & $0$ \\
					$b$ &$\zeta<1$ &  $\nu>\max(\nu_m,\nu_c)$ (H) & p/2 & $2-k(p+2)/4$ & $\frac{p+6}{2}$ & $0$ \\
					\hline
					\label{tbl:Lambda}
				\end{tabular}
			\end{threeparttable}
		}
	\end{table*} 
	
	The angles $\theta_{\rm min}$ and $\theta_{F}$ start decreasing significantly at $t_{\rm{dec}}(\theta_{\rm min,0})$ and $t_{\rm{dec}}(\theta_{F,0})$, respectively.
	In the limit $\theta_{\rm obs}\gg \theta_{F}, \theta_{\rm min}\gg \theta_{\rm c}$ and as long as the flow is still relativistic, both angles decrease as a PL with time.
	Using Eq.~(\ref{eq:thetaminfull}) and approximating $\theta_{\rm obs}-\theta_{\rm min}\approx \theta_{\rm obs}$, we find an
	asymptotic behaviour
	\begin{equation}
	\label{eq:thetamin}
	\theta_{\rm min}
	\approx\theta_{\rm c}\left(\frac{\tilde{t}}{\tilde{t}_{\rm pk}}\right)^\frac{k-3}{a}
	\approx\theta_{\rm obs}\left(\frac{\tilde{t}}{\tilde{t}_{\rm dip}}\right)^\frac{k-3}{a}
	\approx\theta_{*}\left(\frac{\tilde{t}}{\tilde{t}_{*}}\right)^\frac{k-3}{a}
	\propto \tilde{t}^{-(3-k)\over a}\ ,
	\end{equation}
	Where $\tilde{t}_{\rm pk}\approx(\xi_{\rm c} q^2)^{(4-k)/(3-k)}$ is the time of the main peak of the lightcurve (i.e. the latter one, if there are two peaks), $\tilde{t}_{\rm dip}\approx\xi_{\rm c}^{(4-k)/(3-k)}q^{[2(4-k)-a]/(3-k)}$ is the time of the dip in the lightcurve (in case it is double-peaked, see \S \ref{sec:Modelling}) and $\tilde{t}_*\approx\xi_{\rm c}^{[2(4-k)b-a]/[2(b-1)(3-k)]}$ is the deceleration time of the lowest latitude initially beamed at the observer (see \S \ref{sec:Modelling} for more details). As shown in Figures~\ref{fig:thetaminthetaF} and \ref{fig:dFnudOmegamap}, $\theta_{F}(\tilde{t})$ follows a similar asymptotic trend to $\theta_{\rm min}(\tilde{t})$.
	
	\section{lightcurves from angular structure}
	\label{sec:Modelling}
	The angles $\theta_{F}(t),\theta_{\rm min}(t)$ are useful when analyzing the lightcurve of a given burst from a fixed $\theta_{\rm obs}$. In order to qualitatively distinguish between possible lightcurves seen from the same structure, but different observation angles we introduce a beaming angle and time through the relation $\Gamma\theta=1$, i.e. $t_{\rm beam}(\theta)$ is defined through $\theta\Gamma[\theta,t_{\rm beam}(\theta)]\equiv1$, and the corresponding $\theta_{\rm beam}(t)$ is defined by $\theta_{\rm beam}(t)\Gamma[\theta_{\rm beam}(t),t]\equiv1$. Since there could be more than one angle that satisfies this relation, the physically relevant value of $\theta_{\rm beam}(t)$ generally depends on the observation angle. We return to discuss the different regimes in more detail in \S \ref{sec:case1}. 
	
\begin{figure}
		\includegraphics[width=0.45\textwidth]{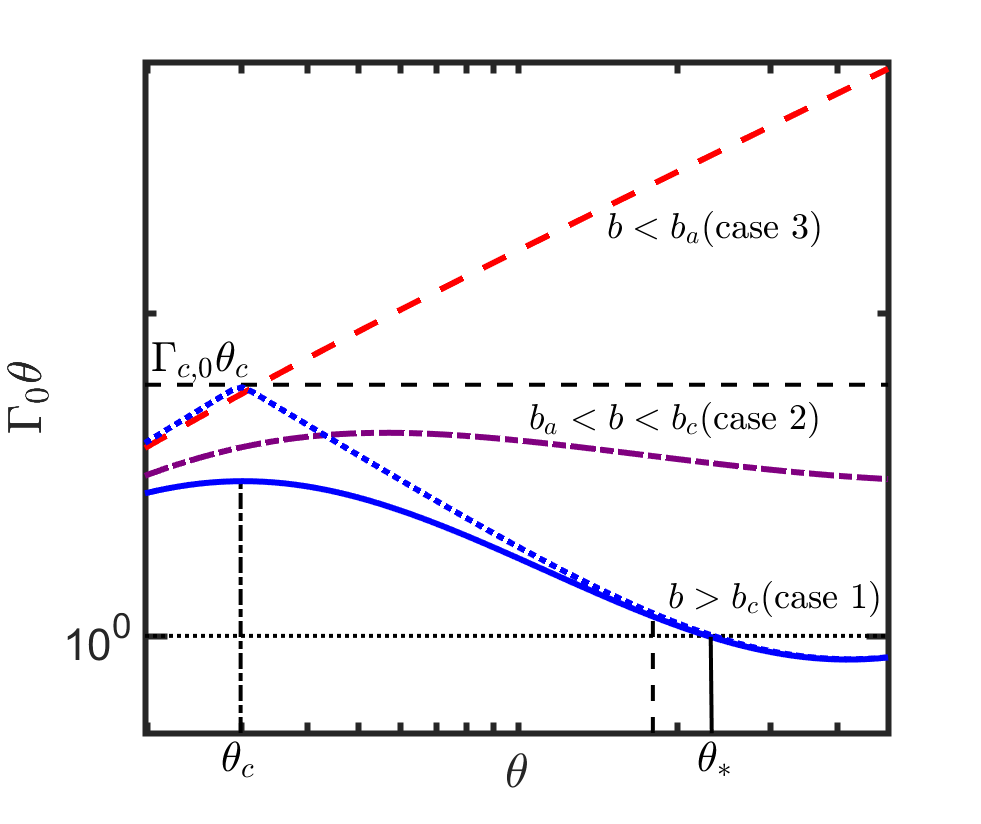}
		\caption{The profile of $\Gamma_0\theta$ for three different values of $b$. We have taken here: $\Gamma_{\rm c}=200,\theta_{\rm c}=0.03$ and $b=2,1.3,0.2$ for cases 1,2,3 respectively. We also consider the case for which $b_c>b_a$ (see \S \ref{sec:Modelling}). A dashed vertical line denotes the approximate solution for $\theta_*$ given by Eq.~(\ref{equation:theta*}) and a solid line depicts the exact value defined by $\Gamma_0(\theta_*)\theta_*=1$. For case 1, we also denote with a dotted blue line, the curve corresponding to the approximation $\Theta \to \max[1,(\theta/\theta_{\rm c})]$. Additional lines depict $\Gamma_{\rm c,0}\theta_{\rm c}$ (dashed horizontal), $\Gamma \theta=1$ (dotted horizontal) and the location of $\theta_{\rm c}$ (dot-dashed vertical).}
		\label{fig:Gammatheta}
	\end{figure}

	Another critical angle is $\theta_*$, which is the value of $\theta$ for which the initial Lorentz factor satisfies $\theta_* \Gamma_0(\theta_*)=1$. It is also approximately the angle for which $\theta_{\rm beam}(t)=\theta_{\rm dec}(t)$, which occurs at the corresponding time $t_*\equiv t_{\rm beam}(\theta_*)\equiv t_{\rm dec}(\theta_*)$ such that $\theta_*=\theta_{\rm beam}(t_*)=\theta_{\rm dec}(t_*)$. This is a critical value, since for $\theta \Gamma_0(\theta)>1$ relativistic beaming from $\theta$ is important from early on and vice versa. For $\theta_*\gg \theta_{\rm c}$, one can approximately write
	\begin{equation}
	\label{equation:theta*}
	\theta_*\approx\theta_{\rm c} \xi_{\rm c}^{1\over 2(b-1)} \quad\Longleftrightarrow\quad q_*\equiv\frac{\theta_*}{\theta_{\rm c}}\approx\xi_{\rm c}^{1\over 2(b-1)}~.
	\end{equation}
	Clearly $b>1$ ($b<1$) is required for a declining (inclining) slope of $\Gamma_0 \theta$ as a function of $\theta$. Furthermore, assuming that\footnote{$\xi_{\rm c}=(\Gamma_{\rm c,0}\theta_{\rm c})^2<1$ is difficult to achieve because of the implied strong lateral causal contact during the acceleration phase, which tends to result in $\Gamma_{\rm c,0}\theta_{\rm c}\gtrsim1$. Moreover, interpreting bright GRBs with a jet break in the afterglow lightcurve as corresponding to $q=\theta_{\rm obs}/\theta_{\rm c}\lesssim1$, afterglow observations suggest $\Gamma_{\rm c,0}\theta_{\rm c}$ of several to a few tens, or $\xi_{\rm c}\sim10^{2}$.} $\xi_{\rm c}>1$, the existence of $\theta_{\rm c}<\theta_*<1$ requires $b>b_c>1$ where
	\begin{equation}
	b_c=-\frac{\log(\Gamma_{\rm c,0})}{\log(\theta_{\rm c})} = 1-\frac{\log(\xi_{\rm c})}{2\log(\theta_{\rm c})}~. 
	\end{equation}
	Another critical value of $b$ is 
	\begin{equation}
	b_a\equiv \frac{a}{2(4-k)}
	\end{equation}
	(separating between jets that decelerate from the core outwards to vice versa). Different physical regimes can arise due to the different possible orderings of $b_a,\,b,\,b_c$. For the purposes of clarity, we assume in what follows that $b_a<b_c$ and explore different values of $b$. This ordering is natural, since for 
	$\Gamma_{\rm c,0}=200, \theta_{\rm c}=0.03$, one finds $b_c\approx 1.5$ (and the value becomes even larger for larger values of either $\Gamma_{\rm c,0}$ or 
	$\theta_{\rm c}$). Therefore, for $k=0$ ($k=2$), $a\gtrsim 12$ ($a\gtrsim 6$) is needed to reverse the condition assumed above. Our division to regimes is thus as follows:
	\begin{description}
		\item 1. $b_a<b_c<b$ with sub-cases: $\theta_{\rm obs}<\theta_*$ (1A) \& $\theta_{\rm obs}>\theta_*$ (1B),\\
		\item 2. $b_a<b<b_c$,\\
		\item 3. $b<b_a<b_c$.
	\end{description}
	The division to the three regimes can be related to the general behaviour of $\Gamma_0 \theta$ as a function of $\theta$ as shown in Fig. \ref{fig:Gammatheta}. We explore below the resulting lightcurves in each of those regimes. We focus on the case  $\Theta \to \max[1,(\theta/\theta_{\rm c})]$, in which the different PL segments can be clearly seen from the figures. 
	We also summarize some of the important distinctions between the different cases in Table~\ref{tbl:cases}.
	
	\begin{table*}
		\centering
		\caption{Summary of lightcurve types in the different cases discussed in this paper.}
		\centering
		\resizebox{0.85\textwidth}{!}{
			\begin{threeparttable}
				\begin{tabular}{cccccc}\hline	
					case &  Condition & $\#$ peaks & timescales  & $\theta_{\rm dec}(t)$ & Approx. $\theta_{\rm min,0}, \theta_{F,0}$\\ \hline
					1A & $\xi_{\rm c}\!>\!1, b\!>\!b_c\!>\!b_a, \theta_{\rm obs}\!<\!\theta_*$ &  2 & $t_{\rm dec}(\theta_{\rm obs})<t_{\rm beam}(\theta_{\rm obs})<t_*$ & increasing & $\theta_{\rm min,0}\approx \theta_{F,0}\approx \theta_{\rm obs}$ \vspace{0.3cm} \\
					1B & $\xi_{\rm c}\!>\!1, b\!>\!b_c\!>\!b_a, \theta_{\rm obs}\!>\!\theta_*$ & 1 & $t_{\rm dec}(\theta_{\rm obs})>t_*$ & increasing & $\theta_{\rm min,0}\approx \theta_{F,0}\ll \theta_{\rm obs}$\vspace{0.3cm} \\
					2 & $\xi_{\rm c}\!>\!1, b_c\!>\!b\!>\!b_a$ &  2 & $t_{\rm dec}(\theta_{\rm obs})<t_{\rm beam}(\theta_{\rm obs})$ & increasing & $\theta_{\rm min,0}\approx \theta_{F,0}\approx \theta_{\rm obs}$ \vspace{0.3cm} \\
					3 & $\xi_{\rm c}\!>\!1, b_c\!>\!b_a\!>\!b$ &  2 & $t_{\rm dec}(\theta_{\rm obs})<t_{\rm beam}(\theta_{\rm obs})$ & declining & $\theta_{\rm min,0}\approx \theta_{F,0}\approx \theta_{\rm obs}$ \vspace{0.3cm}\\
					\hline
					\label{tbl:cases}
				\end{tabular}
			\end{threeparttable}
		}
	\end{table*}

	\subsection{Case 1: $\;\xi_{\rm c}>1,\;\; b>b_c>b_a$}
	\label{sec:case1}
	To understand the expected lightcurve shape, it is useful to consider the orderings of the different timescales in this regime; $\theta_{\rm dec}$ can be expressed as a simple power-law with time, by inverting Eq.~(\ref{eq:t_dec}), 
	\begin{equation}
	\label{eq:thetadec}
	\theta_{\rm dec}(\tilde{t})=\theta_{\rm c} \tilde{t}^{3-k\over 2(4-k)b-a}
	\end{equation}
	where $\tilde{t}\equiv t/t_{\rm dec,c}$. For $\theta_{\rm beam}$, the situation is more subtle. By definition of $\theta_*$, at early times, $\theta_{\rm beam}=\theta_*$. This situation holds until $\tilde{t}=\tilde{t}_{\rm c}=\xi_{\rm c}^{(4-k)/(3-k)}$, which is when the core of the jet has decelerated enough that $\Gamma_{\rm c}\theta_{\rm c}=1$, i.e. $\Gamma_{\rm c}(\tilde{t}_{\rm c})=\theta_{\rm c}^{-1}$ (note that for $b>b_c>b_a$ deceleration occurs first at the core of the jet and only later at the wings). From this moment and until $\tilde{t}_*$, there are two solutions for $\theta_{\rm beam}$, an outer solution at $\theta_{\rm out}=\theta_*$ and an inner solution at $\theta_{\rm in}=\theta_{\rm c}\xi_{\rm c}^{\frac{k-4}{2(4-k)-a}}\tilde{t}^{\frac{3-k}{2(4-k)-a}}$.
	Note that $\theta_{\rm in}(\tilde{t}_{\rm c})=\theta_{\rm c}$, $\theta_{\rm in}(\tilde{t}_{\rm dip})=\theta_{\rm obs}$ and $\theta_{\rm in}(\tilde{t}_*)=\theta_*$ where
	\be
	\label{eq:t*}
	\tilde{t}_*=\tilde{t}_{\rm dec}(\theta_*)=\left(\frac{\theta_*}{\theta_{\rm c}}\right)^\frac{2(4-k)b-a}{3-k}=\xi_{\rm c}^\frac{2(4-k)b-a}{2(b-1)(3-k)}\ .
	\ee
	The situation changes once more at $\tilde{t}_*$, since beyond this time, the entire jet satisfies $\Gamma \theta<1$ and therefore $\theta_{\rm beam}$ is no longer defined.
	The physically relevant $\theta_{\rm beam}(t)$ depends on $\theta_{\rm obs}$. For $\theta_{\rm obs}>\theta_*$, $\theta_{\rm in}$ never dominates the lightcurve. This is because by the time material from this angle is beamed towards the observer, $\theta_{\rm beam}$ is no longer defined (see above). Even $\theta_{\rm out}=\theta_*$ is physically significant only for $\theta_{\rm obs}\approx\theta_*$. Alternatively, for $\theta_{\rm obs}<\theta_*$, $\theta_{\rm out}=\theta_*$ carries less energy than material travelling along the line of sight and is never physically important.
	We conclude that
	\begin{eqnarray}
	\label{eq:thetabeam}
	\theta_{\rm beam}(\tilde{t})=\left\{ \!\begin{array}{ll}\!\theta_{\rm out}=\theta_* & \theta_{\rm obs}>\theta_*\,,\ \ \tilde{t}< \tilde{t}_{*}\,,\\
	\!\theta_{\rm in}\!=\!\theta_{\rm c}\!\fracb{\tilde{t}^{3-k}}{\xi_{\rm c}^{4-k}}^{\frac{1}{2(4-k)-a}}
	& \theta_{\rm obs}\!<\!\theta_*\,,\ \ \tilde{t}_{\rm c}\!<\!\tilde{t}\!<\!\tilde{t}_*\,,
	\end{array} \right.
	\end{eqnarray}
	In particular, for $\tilde{t}_{\rm c}<\tilde{t}<\tilde{t}_*$ there are three regions in terms of the relationship between $\theta_{\rm obs},\theta_{\rm in}, \theta_{\rm out}$ which are divided as follows:
	\begin{enumerate}
		\item  $\theta_{\rm obs} < \theta_{\rm in}(t)$  -- Here $t>t_{\rm beam}(\theta_{\rm obs})$ which leads to $\theta_{F} \approx \theta_{\rm min} \ll \theta_{\rm obs}$. As we will show below this corresponds to the shallow rising phase of the lightcurve, $F_{\nu}\propto t^{\alpha}$.
		\smallskip
		\item $\theta_{\rm in}(t) < \theta_{\rm obs} < \theta_{\rm out}=\theta_*$ -- Here $\theta_{F} \approx \theta_{\rm min} \approx \theta_{\rm obs}$. This represents the early (first inclining and then declining) part of the lightcurve, as will be detailed below.
		\smallskip
		\item $\theta_{\rm obs} > \theta_{\rm out}=\theta_*$ -- Here $\theta_{F,0} \approx \theta_{\rm min,0} \,\approx \theta_*(\theta_{\rm obs}/\theta_{*})^{1/b} \ll \theta_{\rm obs}$. In this case, the behaviour changes after $t_{\rm dec}(\theta_{F,0})\approx t_*(\theta_{\rm obs}/\theta_{*})^{[2(4-k)b-a]/(3-k)b}>t_*$, since this is when $\theta_{F}, \theta_{\rm min}$ start to decrease significantly and the shallow rising part of the lightcurve, $F_{\nu}\propto t^{\alpha}$, emerges.
	\end{enumerate}
	
\begin{figure}
		\includegraphics[width=0.45\textwidth,height=0.3\textwidth]{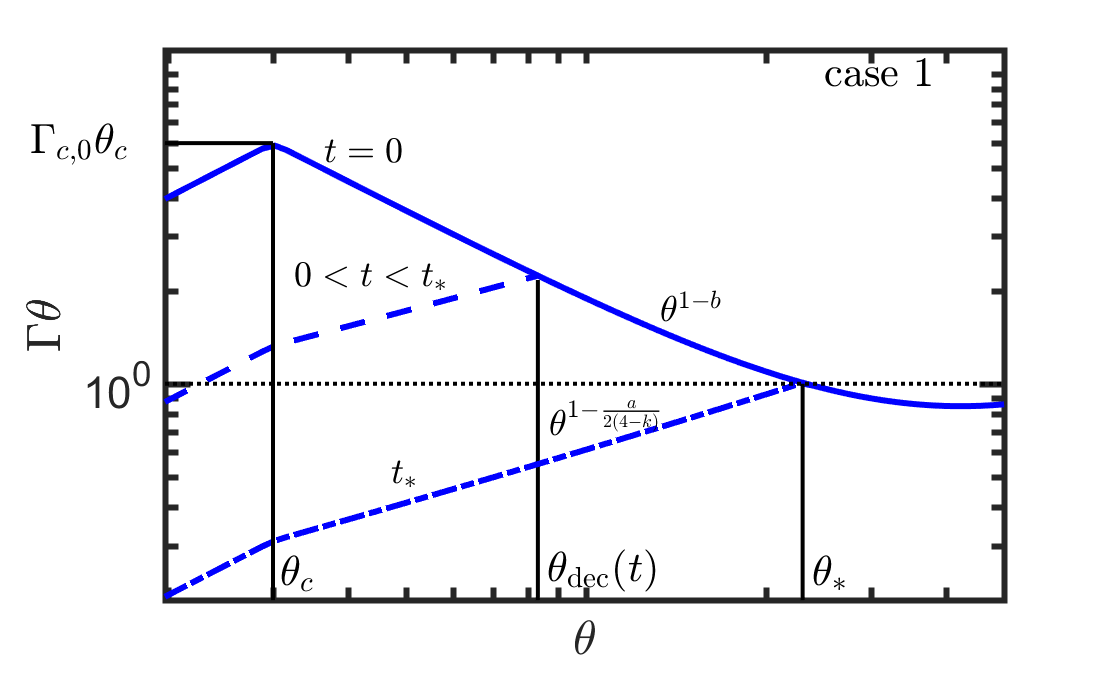}
		\includegraphics[width=0.45\textwidth,height=0.3\textwidth]{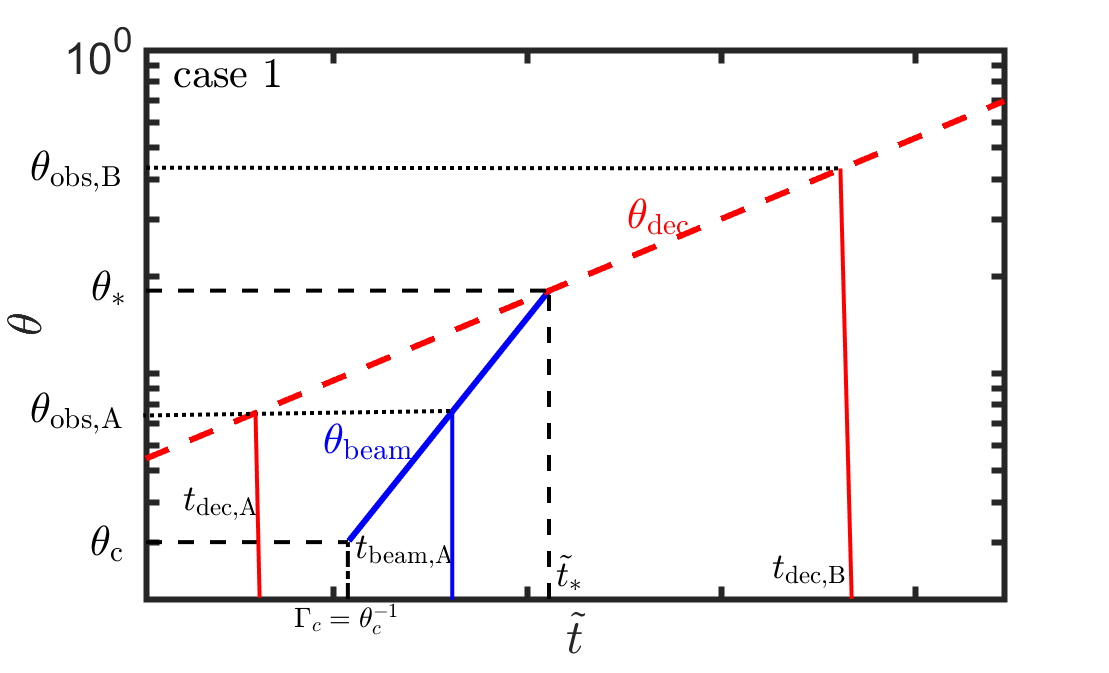}
		\caption{\textbfit{Top}: Temporal evolution of $\Gamma \theta$ in case 1 ($b>b_c$). Results are plotted for $\theta_{\rm c}=0.03,\Gamma_{\rm c}=200$ as well as $k=0,a=4,b=2$. \textbfit{Bottom}: Corresponding evolution of the characteristic angles with time. For $\theta_{\rm obs}=\theta_{\rm obs,A}<\theta_*$, one gets $t_{\rm dec,A}<t_{\rm beam,A}<t_*$ while for $\theta_{\rm obs}=\theta_{\rm obs,B}>\theta_*$, one gets $t_{\rm dec,B}>t_*$ (and $t_{\rm beam}$ becomes non defined in this case).} 
		\label{fig:A1theta}
	\end{figure}

\begin{figure}
		\includegraphics[width=0.45\textwidth]{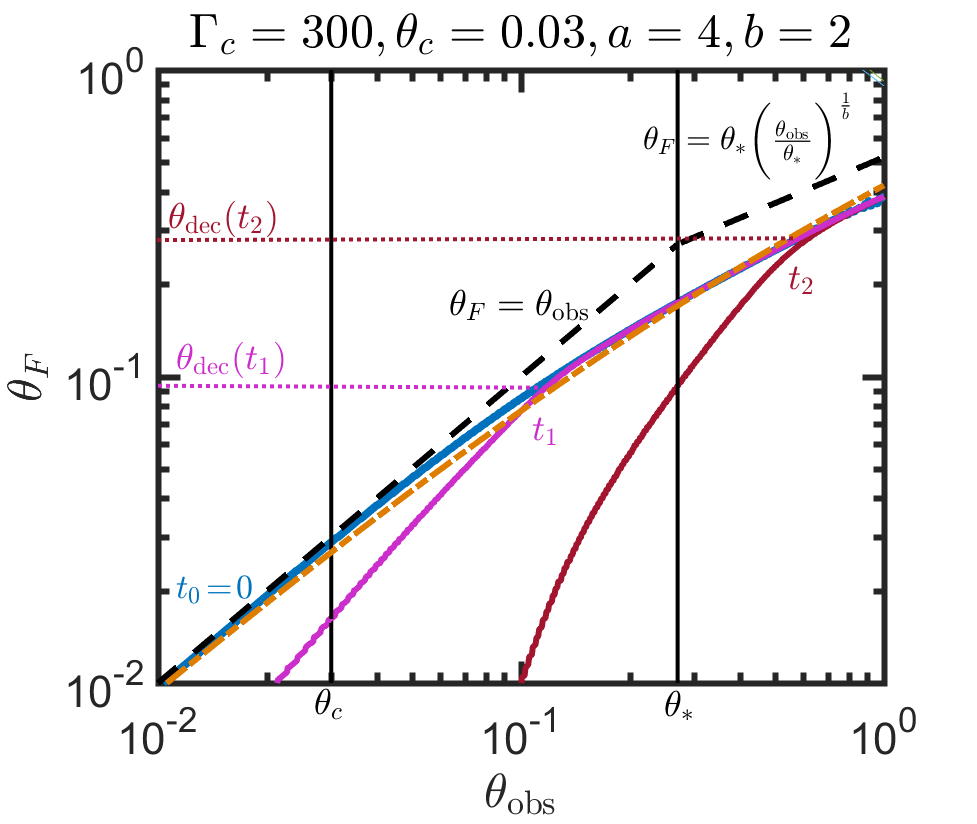}
		\caption{$\theta_{F}(\theta_{\rm obs})$ for different observation times ($0=t_0<t_1<t_2$). Results are plotted for $\theta_{\rm c}=0.03,\Gamma_{\rm c}=300$ as well as $k=0,a=4,b=2$. As a comparison we also plot in a dot-dashed line the approximate value of $\theta_{F,0}$ given by equation \ref{eq:thetaF0full}, as well as the asymptotic scalings $\theta_{F,0}\approx\theta_{\rm obs}$ and $\theta_{F,0}\approx 
			\theta_*(\theta_{\rm obs}/\theta_{*})^{1/b}$ expected to hold for $\theta_{\rm obs}\ll \theta_*$ and $\theta_{\rm obs}\gg \theta_*$ respectively in dashed lines. Horizontal dotted lines depict $\theta_{\rm dec}(t_i)$ for each case as given by equation \ref{eq:thetadec}.}
		\label{fig:thetaFthetaobscomb}
	\end{figure}

	The situation is demonstrated by observing the temporal evolution of $\Gamma \theta$ and the direct evolution of $\theta_{\rm beam},\theta_{\rm dec}$ in Fig.~\ref{fig:A1theta}.
	Evidently, two sub-cases exist here depending on $\theta_{\rm obs}/\theta_*$. We explore those sub-cases below.
	We also present the evolution of $\theta_{F}(\theta_{\rm obs})$ as a function of time in Fig.~\ref{fig:thetaFthetaobscomb}. This figure demonstrates the validity of our approximation for $\theta_{F,0}$, given by Eq.~(\ref{eq:thetaF0full}). For $\theta_{\rm obs}\ll\theta_*$, $\theta_{F,0}\to \theta_{\rm obs}$ as expected. For $\theta_{\rm obs}\gg\theta_*$ we have $\theta_{F,0}\approx\theta_{\rm min,0}\approx  \theta_{*} (\theta_{\rm obs}/\theta_*)^{1/b}$. However, in practice, for finite values of $\theta_{\rm obs}/\theta_{*}$, the real value of $\theta_{F,0}$ is slightly below the above approximation. Furthermore, since the core decelerates faster than the wings, the further $\theta_{\rm obs}$ is from $\theta_{\rm c}$, the longer it takes for $\theta_{F}$ to start diminishing significantly.

	\begin{enumerate}
		\item {\bf case 1A, $\theta_{\rm obs}<\theta_*$:} Here the ordering of the timescales is $t_{\rm dec}(\theta_{\rm obs})<t_{\rm beam}(\theta_{\rm obs})<t_*$. In this case, the emission from $\theta<\theta_{\rm obs}$ is initially strongly beamed away from the observer, while the material along the line of sight whose emission is beamed towards the observer lies within $|\theta-\theta_{\rm obs}|\lesssim1/\Gamma(\theta_{\rm obs})\ll\theta_{\rm obs}$, so the early emission is dominated by material near $\theta_{F,0}\approx\theta_{\rm min,0}\approx\theta_{\rm obs}$.
		For $t<t_{\rm dec}(\theta_{F,0})$ the flux therefore rises as $F_{\nu}\propto t^{\alpha_r}$ while the line of sight material hasn't yet decelerated (where $\alpha_r$ depends on $k$ and on the observed PLS, see tables \ref{tbl:Lambda}, \ref{tbl:alpha}). After a short, intermediate duration between $t_{\rm dec}(\theta_{F,0}), t_{\rm dec}(\theta_{\rm obs})$, the local dynamics of the material along the LOS begin to follow a largely spherical self-similar evolution \citep{BM76} and the resulting flux is similar to cosmological GRBs viewed on-axis $F_{\nu}\propto t^{\alpha_d}$ (where for example, for $k=0$ and PLS G $\alpha_d= 3(1-p)/4$, see tables \ref{tbl:Lambda}, \ref{tbl:alpha}). In practice the decay of the lightcurve is not as steep as in the spherical case, due to the fact that material from $\theta>\theta_{\rm obs}$ decelerates after $t_{\rm dec}(\theta_{\rm obs})$, and its contributions cannot be completely ignored (see Fig. \ref{fig:dFnudOmegamap}).
		This phase lasts until $\tilde{t}_{\rm dip}\approx \tilde{t}_{\rm beam}(\theta_{\rm obs})\approx q^{2(4-k)-a\over 3-k}\xi_{\rm c}^{4-k \over 3-k}$ which is approximately the time when material internal to the line of sight starts becoming visible as its beaming cone reaches the observer. At $t_{\rm dip}<t<t_{\rm pk}$, the emission becomes dominated by material at progressively smaller $\theta\sim\theta_{\rm min}\sim \theta_{F}\ll\theta_{\rm obs}$ (see Eq.~(\ref{eq:thetamin})). This typically leads to a shallow rise in the flux, $F_{\nu}\propto t^{\alpha}$ (see Appendix A2 of GG18 for a derivation of the asymptotic $\alpha$ in this phase and Table~\ref{tbl:alpha} for the values corresponding to different PLS). The rise continues until $t_{\rm pk}$, when the jet's core becomes visible (i.e. $\theta_{\rm min}(t_{\rm pk})\to 0$; see \S \ref{sec:Inferences}). Beyond this point, the full jet becomes visible to the observer and the lightcurve evolves as for an on-axis GRB jet post jet-break, $F_{\nu}\propto t^{\alpha_f}$ \cite[for a detailed discussion of this phase see e.g.][]{Granot07,Granot-Piran-12,DeColle+12b,Gill+19}.
		To calculate the analytic lightcurve, the flux of each peak is calculated by
		\begin{equation}
		\label{eq:pkstruct}
		F=\bar{F}\, 2^{\alpha_1-\alpha_2 \over 2}\, \bar{t}^{\alpha_1}\, (1+\bar{t}^2)^{\alpha_2-\alpha_1\over 2}
		\end{equation}
		where $\bar{t}$ is the time normalized to the peak time, $\bar{F}$ is the peak flux and $\alpha_1, \alpha_2$ are the temporal slopes before and after the peak respectively.
		The overall flux is a sum of two terms of the form given by Eq.~(\ref{eq:pkstruct}) for the two peaks, i.e.
		\begin{eqnarray}
		& F=F_{\rm 1pk}2^{\alpha_r-\alpha_d \over 2}\bigg(\frac{t}{t_{\rm 1pk}}\bigg)^{\alpha_r}\bigg[1+\bigg(\frac{t}{t_{\rm 1pk}}\bigg)^2\bigg]^{\alpha_d-\alpha_r\over 2} \nonumber  \\ & + F_{\rm pk}2^{\alpha-\alpha_f \over 2}\bigg(\frac{t}{t_{\rm pk}}\bigg)^{\alpha}\bigg[1+\bigg(\frac{t}{t_{\rm pk}}\bigg)^2\bigg]^{\alpha_f-\alpha\over 2}
		\end{eqnarray}
		where $t_{\rm 1pk}, t_{\rm pk}, F_{\rm 1pk}, F_{\rm pk}$ are correspondingly the times and fluxes of the first and second fluxes. A summary of their values is given in \S \ref{sec:Inferences}.
		An illustration of the overall lightcurve in this case is shown in Fig. \ref{fig:1Alight}, side by side with the result of the numerical calculation of GG18. The analytic prescription provides a good approximation of the more complete calculation.

		\begin{figure}
			\includegraphics[width=0.45\textwidth]{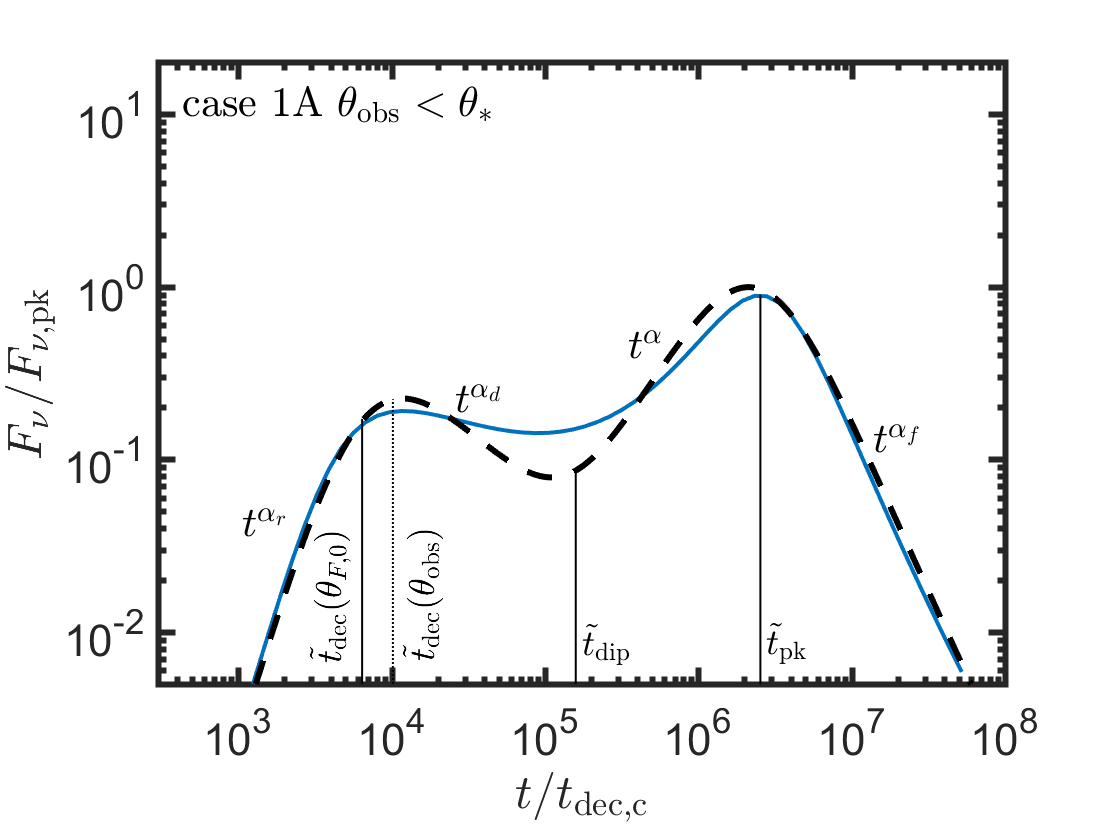}\\
			\caption{Analytic lightcurve obtained for case 1A (dashed) as compared with the numerical model of GG18 (solid). Results are shown for $k=0,p=2.2,a=4,b=2,\Gamma_{\rm c}=1000,\theta_{\rm c}=0.03,\theta_{\rm obs}=0.3$ and PLS G. $t_{\rm dec}(\theta_{F,0})$ is calculated using Eq.~(\ref{eq:thetaFprox}).}
			\label{fig:1Alight}
		\end{figure}

		\item {\bf  case 1B, $\theta_{\rm obs}>\theta_*$:} Here the ordering of timescales is $t_*<t_{\rm dec}(\theta_{\rm obs})$, which implies that a wide range of angles $\theta$ such that $\theta_{\rm obs}\gg\theta\gtrsim\theta_{F,0}$ are visible from the very start. 
		As a first approximation, it is constructive to consider the approximation $\Gamma_0^{-1}(\theta_{\rm min,0})=\theta_{\rm obs}-\theta_{\rm min,0}\approx\theta_{\rm obs}$ leading to $\theta_{F,0}\approx \theta_{\rm min,0}\approx \theta_{\rm c}(\xi_{\rm c}q^2)^{1/2b} \approx\theta_*(\theta_{\rm obs}/\theta_{*})^{1/b}\ll \theta_{\rm obs}$,
		where the time of significant decrease in $\theta_F\approx\theta_{\rm min}$ is expected to be around $t_{\rm dec}(\theta_{F,0})\approx t_{\rm dec}(\theta_{\rm min,0})
		\approx(\xi_{\rm c}q^2)^{[2(4-k)b-a]/[2b(3-k)]}
		\approx t_*(\theta_{\rm obs}/\theta_{*})^{[2(4-k)b-a]/(3-k)b}>t_*$.  
		However, as shown in Fig.~\ref{fig:thetaFthetaobscomb}, this approximation is valid only for $\theta_{\rm obs}\gg \theta_*$ and in practice it somewhat overestimates $\theta_{F,0}$, and correspondingly $t_{\rm dec}(\theta_{F,0})$. For a more accurate approximation we therefore apply Eq.~(\ref{eq:thetaFprox}). Alternatively, one can more conveniently use the analytic approximation in Eq.~(\ref{eq:thetaF0full}).

		In general, the most significant contributions to the emission come from material that has both decelerated and whose emission is beamed towards the observer. At $t=t_{\rm dec}(\theta_{F,0})$ the material from $\theta_{F,0}$ satisfies both these conditions, and indeed it dominates the observed emission at that time.
		At earlier times the material at $\theta_{F,0}$, had the same $\Gamma$ as at deceleration and was therefore still beamed towards the observer, although not yet slowed down. The result is that for $t<t_{\rm dec}(\theta_{F,0})$ the flux is still dominated by material at $\theta_{F,0}$ and rises as $F_{\nu}\propto t^{\alpha_r}$, where $\alpha_r$ is the pre-deceleration rise of the line-of-sight flux and depends on the spectral regime that is observed (see Table~\ref{tbl:Lambda}). At later times $t_{\rm dec}(\theta_{F,0})<t<t_{\rm pk}$ the flux becomes dominated by $\theta\sim\theta_{F}(t)\sim\theta_{\rm min}(t)$ (see Eq.~(\ref{eq:thetamin})) and evolves as $F_{\nu}\propto t^{\alpha}$ as described in case 1A above. The flux at $t>t_{\rm pk}$ evolves (as for 1A) according to the standard post jet break scaling. An illustration of the lightcurve obtained in this case is given in Fig. \ref{fig:1Blight} alongside the numerical calculation of GG18. Note that emission from material along the line of sight is always sub-dominant in this case. An expression for the flux is given by
		\begin{eqnarray}
		F\!=\!2^{\alpha\!-\!\alpha_f \over 2}F_{\rm pk} \bigg[1\!+\!\bigg(\frac{t}{t_{\rm dec}(\theta_{F,0})}\bigg)^{-4}\bigg]^{\alpha\!-\!\alpha_r\over 4} \bigg(\frac{t}{t_{\rm pk}}\bigg)^{\alpha}\bigg[1\!+\!\bigg(\frac{t}{t_{\rm pk}}\bigg)^2\bigg]^{\alpha_f\!-\!\alpha\over 2}
		\end{eqnarray}

		\begin{figure}
			\includegraphics[width=0.45\textwidth]{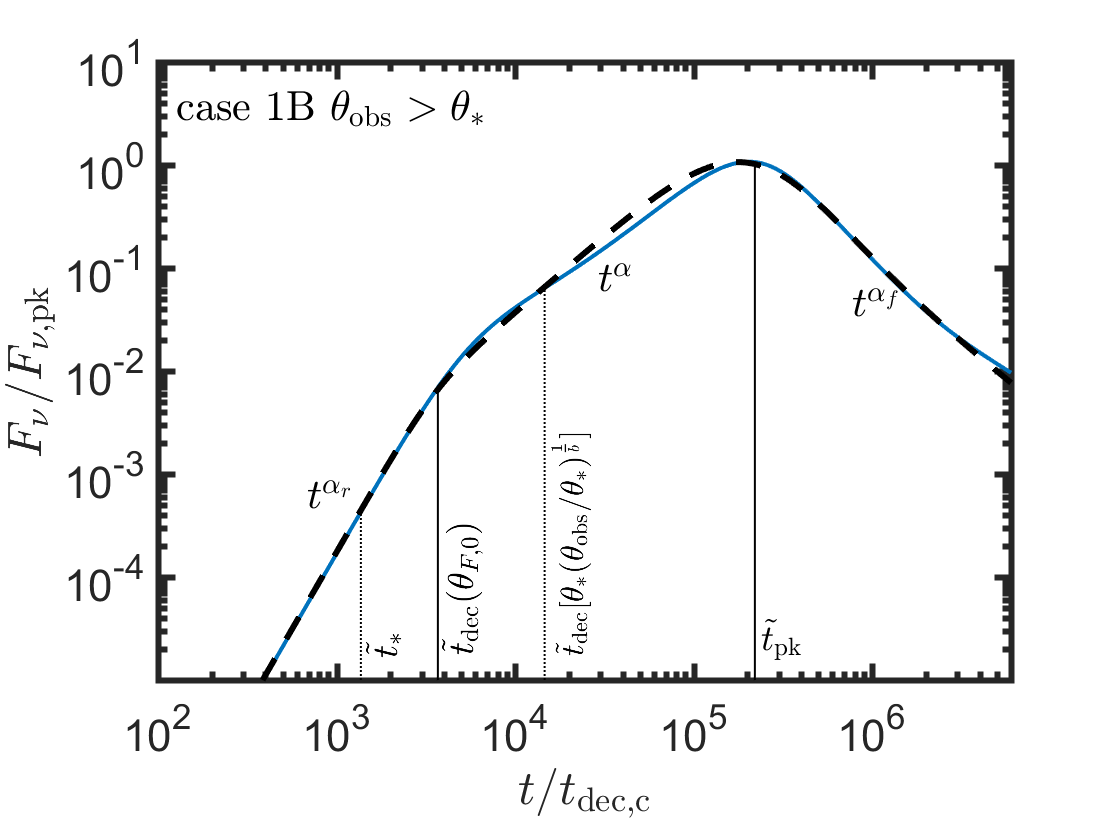}\\
			\caption{Analytic lightcurve obtained for case 1B (dashed) as compared with the numerical model of GG18 (solid). Results are shown for $k=0,p=2.2,a=4,b=2,\Gamma_{\rm c}=200,\theta_{\rm c}=0.03,\theta_{\rm obs}=0.6$ and PLS G; $t_{\rm dec}(\theta_{F,0})$ is calculated using Eq.~(\ref{eq:thetaFprox}).}
			\label{fig:1Blight}
		\end{figure}
		
	\end{enumerate}
	
	\subsection{Case 2: $\;\xi_{\rm c}>1,\;\; b_a<b<b_c$}
\begin{figure}
		\includegraphics[width=0.24\textwidth]{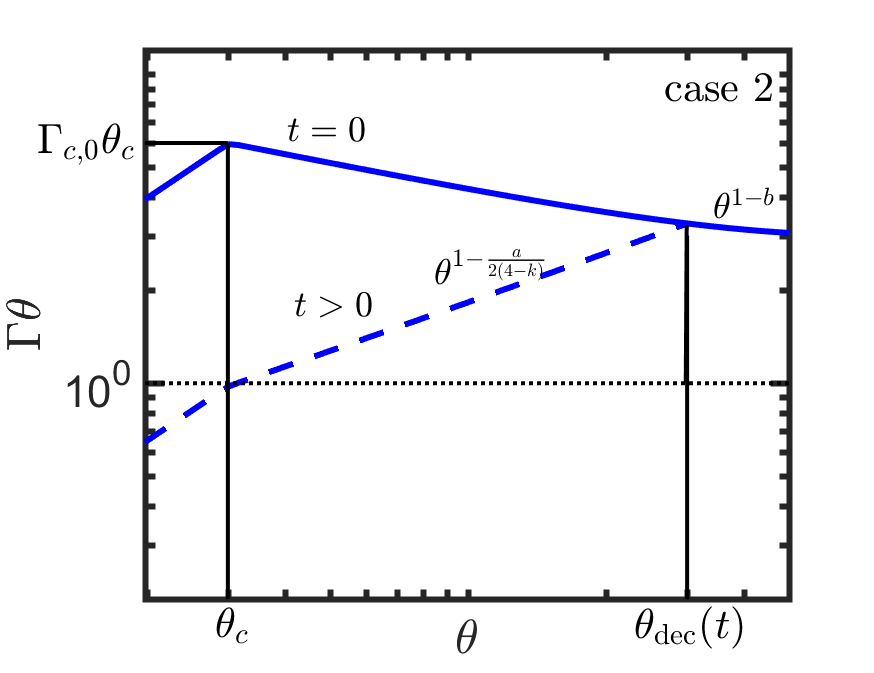}
		\includegraphics[width=0.24\textwidth]{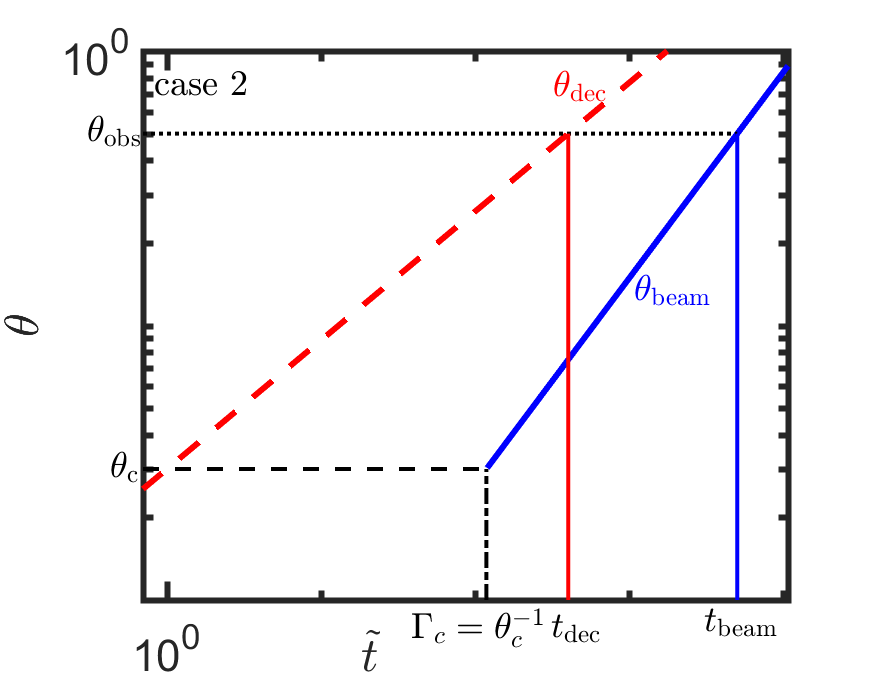}\\
		\caption{Left: Temporal evolution of $\Gamma \theta$ in case 2 ($b_a<b<b_c$). Results are plotted for $\theta_{\rm c}=0.03,\Gamma_{\rm c}=200$ as well as $k=0,a=4,b=1.3$. Right: Corresponding evolution of the characteristic angles with time. $t_{\rm dec}<t_{\rm beam}$ for any observation angle.} 
		\label{fig:2}
	\end{figure}

\begin{figure}
		\includegraphics[width=0.45\textwidth]{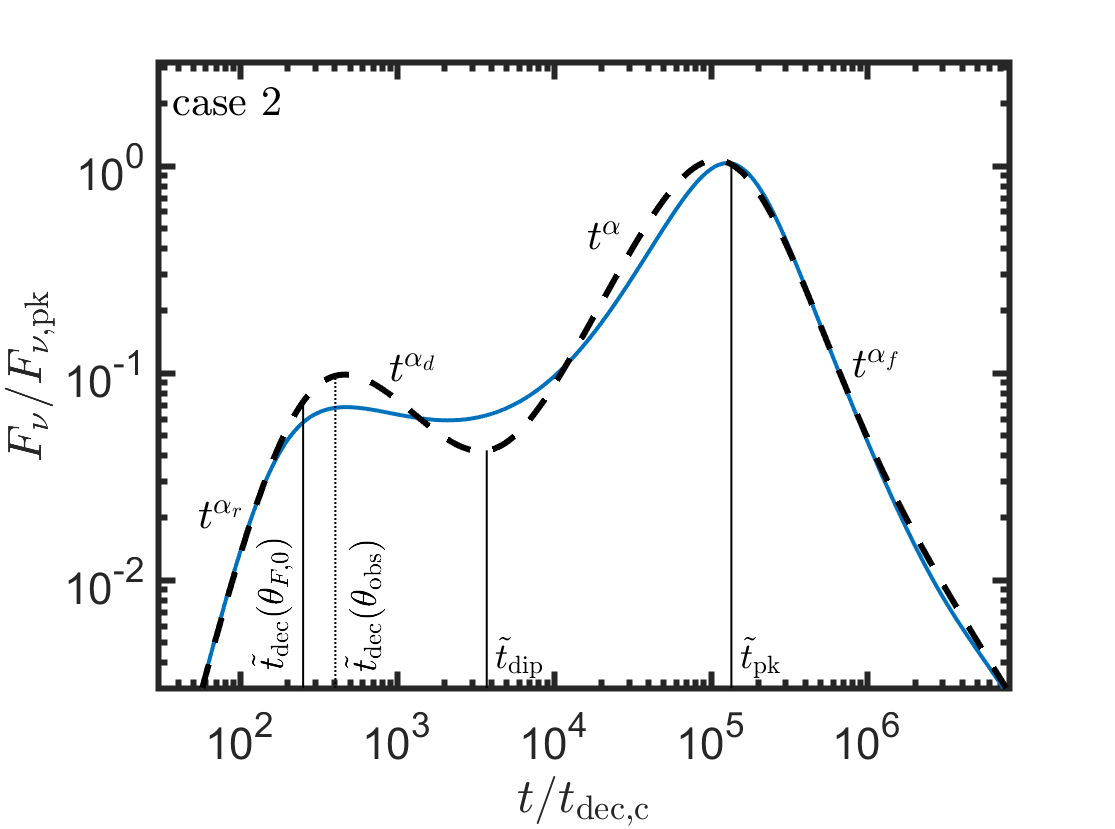}
		\caption{Analytic lightcurve obtained for case 2 (dashed) as compared with the numerical model of GG18 (solid). Results are shown for $k=0,p=2.2,a=4,b=1.3,\Gamma_{\rm c}=200,\theta_{\rm c}=0.03,\theta_{\rm obs}=0.5$ and PLS G. $t_{\rm dec}(\theta_{F,0})$ is calculated using Eq.~(\ref{eq:thetaFprox}).} 
		\label{fig:2light}
	\end{figure}

	The expressions for $\theta_{\rm dec}$ and $\theta_{\rm beam}$ remain the same in this case as in case 1 above and are given by Eqs.~(\ref{eq:thetadec}) and (\ref{eq:thetabeam}), respectively. The difference here is that $\theta_*$ is no longer defined since $\theta\Gamma_0(\theta)>1$ for all $\theta$. The resulting situation is equivalent to case 1A (i.e. $b>b_c$ with $\theta_{\rm obs}<\theta_*$) in which $t_{\rm dec}(\theta_{\rm obs})<t_{\rm beam}(\theta_{\rm obs})$). This is demonstrated in Fig. \ref{fig:2} where we plot the temporal evolution of $\Gamma\theta$ and of the critical angles.
	We also plot the resulting lightcurve in Fig. \ref{fig:2light}.

	\subsection{Case 3: $\;\xi_{\rm c}>1,\;\; b<b_a<b_c$}
\begin{figure}
		\includegraphics[width=0.24\textwidth]{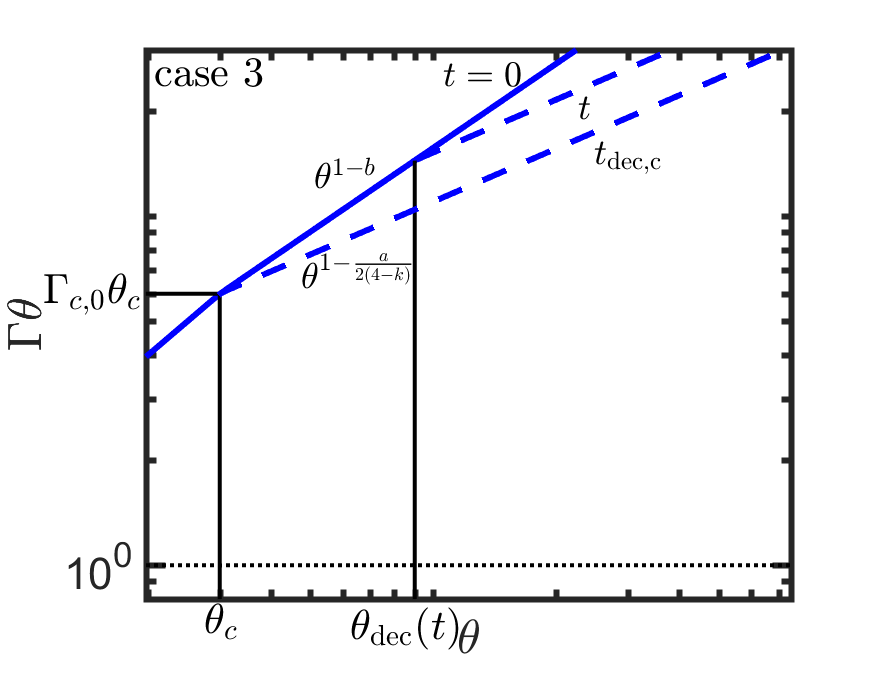}
		\includegraphics[width=0.24\textwidth]{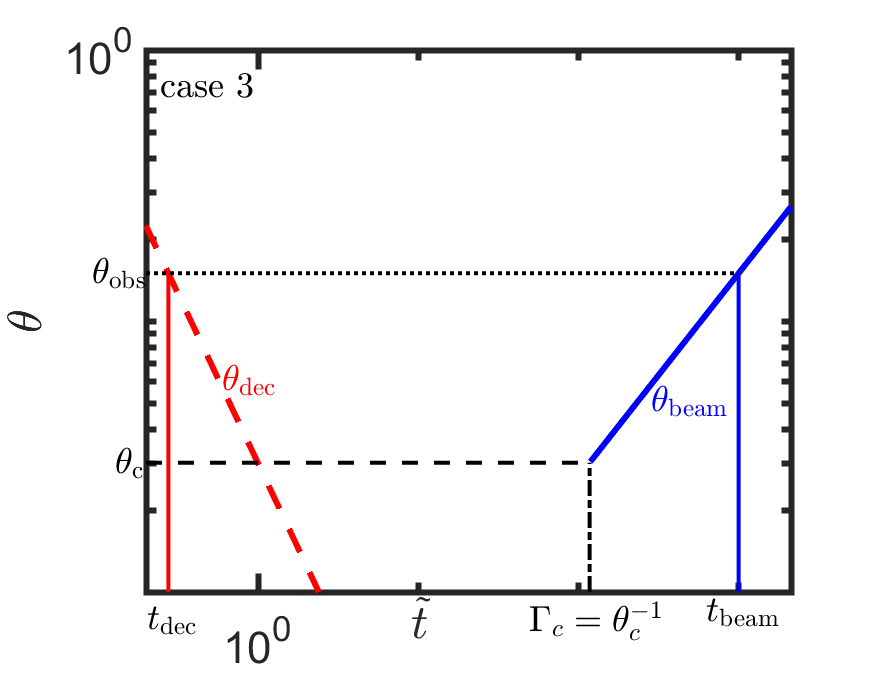}\\
		\caption{Left: Temporal evolution of $\Gamma \theta$ in case 3 ($b<b_a<b_c$). Results are plotted for $\theta_{\rm c}=0.03,\Gamma_{\rm c}=200$ as well as $k=0,a=4,b=0.2$. Right: Corresponding evolution of the characteristic angles with time. $t_{\rm dec}<t_{\rm beam}$ for any observation angle.} 
		\label{fig:3}
	\end{figure}

\begin{figure}
		\includegraphics[width=0.45\textwidth]{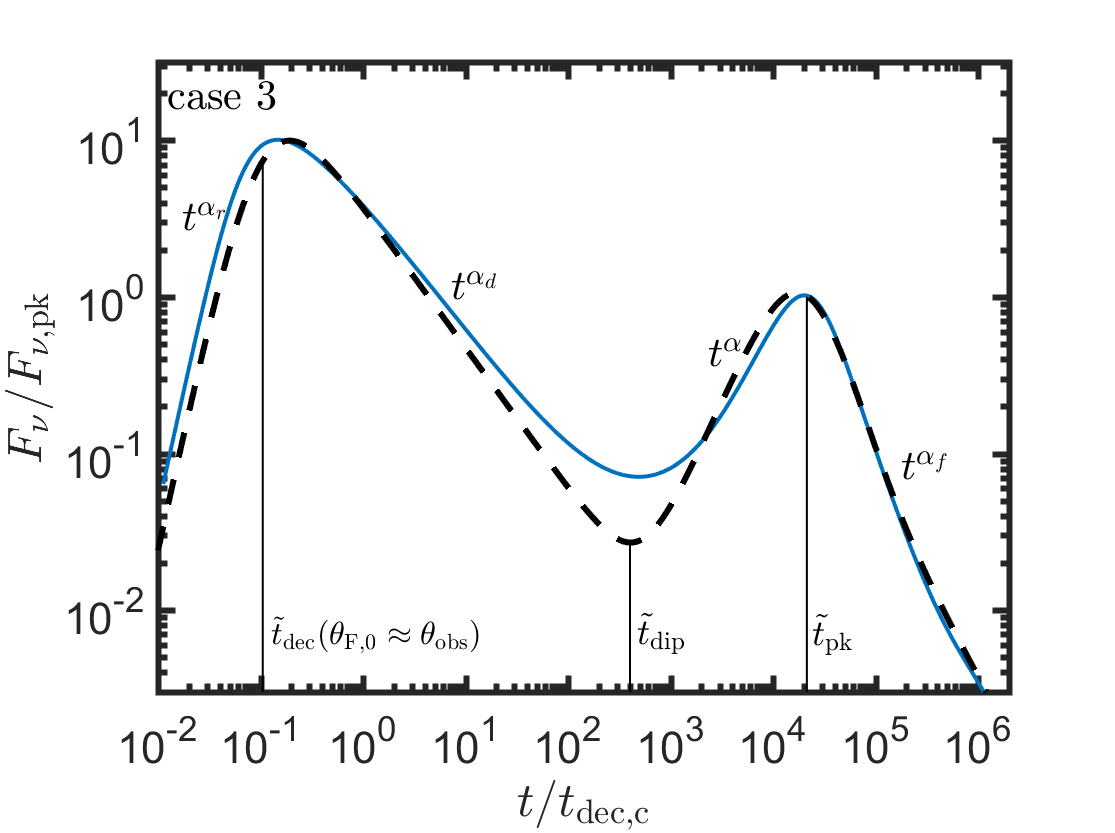}
		\caption{Analytic lightcurve obtained for case 3 (dashed) as compared with the numerical model of GG18 (solid). Results are shown for $k=0,p=2.2,a=4,b=0.2,\Gamma_{\rm c}=100,\theta_{\rm c}=0.03,\theta_{\rm obs}=0.5$ and PLS G.} 
		\label{fig:3light}
	\end{figure}

	In this regime the wings of the jet are sufficiently fast, that deceleration progresses from the outside in rather than vice versa as in the previous cases. As in case 2, since $\Gamma_0\theta>1$ for any $\theta$, $\theta_*$ is not defined in this case. Once more, the resulting evolution is similar to cases 1A and 2, i.e. $t_{\rm dec}(\theta_{\rm obs})<t_{\rm beam}(\theta_{\rm obs})$. This is shown in Fig. \ref{fig:3} where we plot the temporal evolution of $\Gamma\theta$ and the critical angles. We also plot the resulting lightcurve in Fig. \ref{fig:3light}.

	\subsection{Gaussian jets}
	\label{sec:Gauss}
\begin{figure}
		\includegraphics[width=0.24\textwidth]{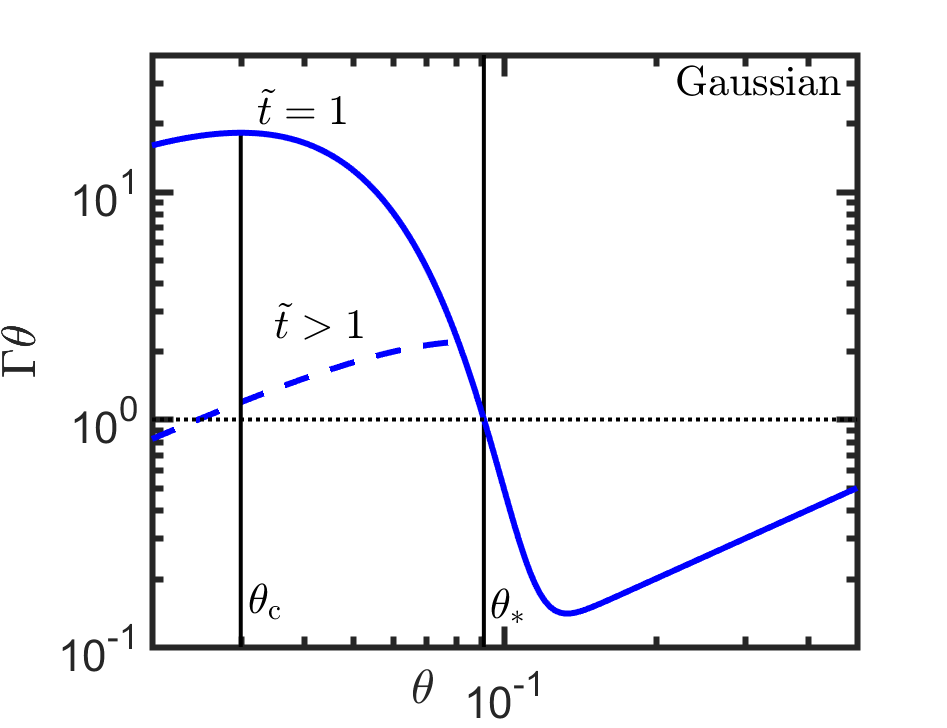}
		\includegraphics[width=0.24\textwidth]{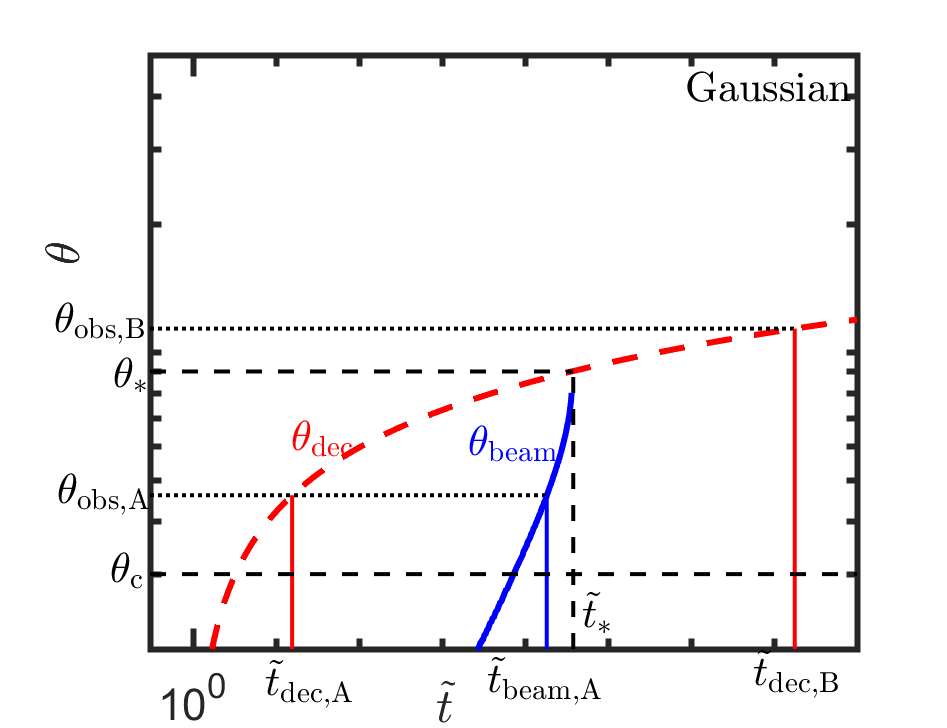}
		\caption{Left: Temporal evolution of $\Gamma \theta$ for a Gaussian structured jet. Results are plotted for $\theta_{\rm c}=0.03,\Gamma_{\rm c}=1000$ as well as $k=0$. Right: Corresponding evolution of the characteristic angles with time.} 
		\label{fig:Gauss}
	\end{figure}

\begin{figure}
		\includegraphics[width=0.24\textwidth]{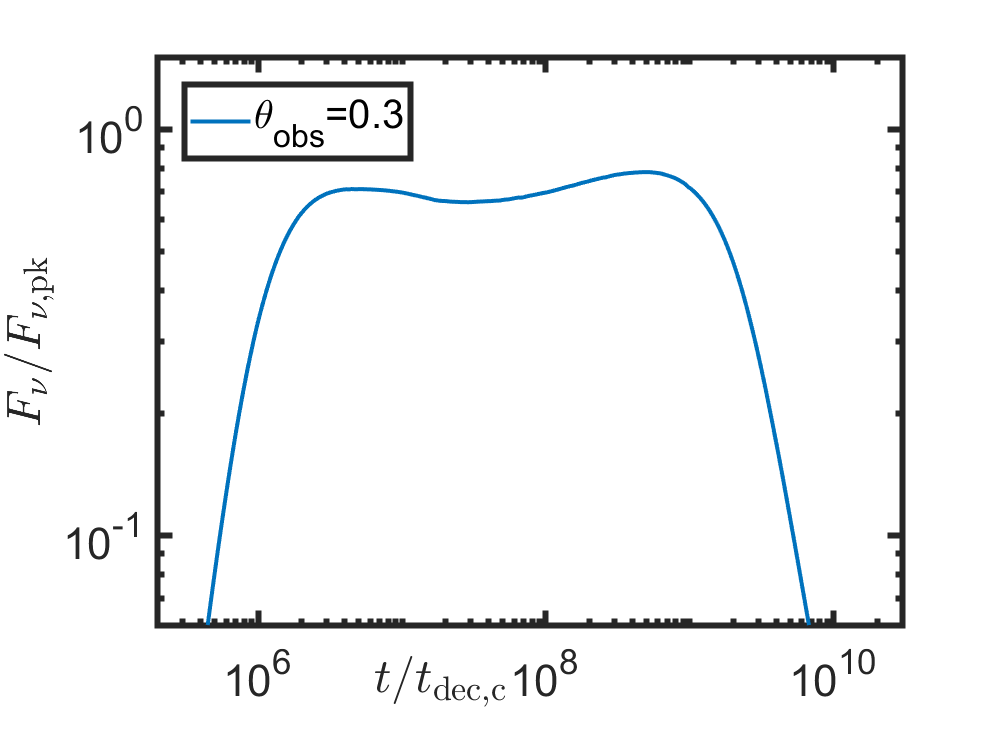}
		\includegraphics[width=0.24\textwidth]{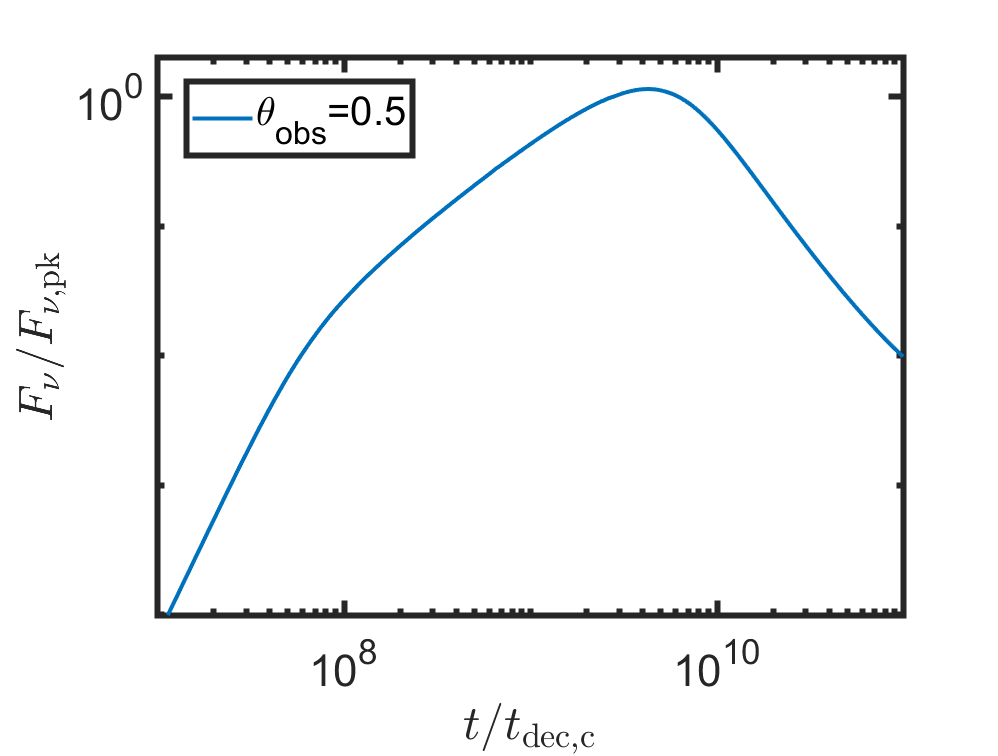}
		\caption{Representative lightcurves calculated using the numerical model of GG18 applied to Gaussian structured jets. The curves correspond to $\theta_{\rm obs}<\theta_*$ (left) and $\theta_{\rm obs}>\theta_*$ (right). Results are plotted for $\Gamma_{\rm c,0}=10^4, \theta_{\rm c}=0.08$ and $\theta_{\rm obs}=0.3,0.5$ respectively.} 
		\label{fig:Gausslc}
	\end{figure}

	We have focused so far on PL jets, for which closed expressions for, e.g. $\theta_{\rm beam}(t)$ can be obtained. For completeness, we briefly discuss here the case of Gaussian structures for the energy and Lorentz factor  (e.g. \citealt{Rossi02,Zhang2002,Kumar-Granot-03}):
	\begin{equation}\label{eq:PLgauss}
	\frac{\epsilon}{\epsilon_{\rm c}} = e^{-\theta^2/2\theta_{\rm c}^2}~,\quad\frac{\Gamma_0(\theta)-1}{\Gamma_{\rm c,0}-1} =  e^{-\theta^2/2\theta_{\rm c}^2}~\ ,
	\end{equation}
	Following the same derivation outlined in \S \ref{sec:definitions}, one obtains
	\begin{equation}
	\tilde{t}_{\rm dec}(\theta)=e^{\frac{\theta^2}{2\theta_{\rm c}^2}\frac{7-2k}{3-k}}
	\end{equation}
	For $k=0$, $\tilde{t}_{\rm dec}(\theta)$ increases with $\theta$. In addition, since $\Gamma(\theta)$ decreases quickly, $\theta_*$, defined by the implicit equation
	\begin{equation}\label{eq:Gaussian_qs}
	\xi_{\rm c} q_*^2 e^{-q_*^2}=1\quad,\quad q_*=\theta_*/\theta_{\rm c}\ ,
	\end{equation}
	typically satisfies $\theta_*<1$.
	The result is that the Gaussian case is qualitatively similar to case 1 (\S \ref{sec:case1}), with equivalent A and B sub-cases. Namely, if $\theta_{\rm obs}<\theta_*$, then $t_{\rm dec}(\theta_{\rm obs})<t_{\rm beam}(\theta_{\rm obs})$, resulting in a double peaked lightcurve, and if $\theta_{\rm obs}>\theta_*$, then $t_{\rm dec}(\theta_{\rm obs})>t_{\rm beam}(\theta_{\rm obs})$, resulting in a single peaked lightcurve. These results are depicted in Figures~\ref{fig:Gauss}, \ref{fig:Gausslc}. Note that a large $\xi_{\rm c}$ is required in order to have $\theta_*\gg \theta_{\rm c}$ as required in order for the double-peaked lightcurve to be realized in practice.

\begin{figure}
		\centering
		\includegraphics[width=0.44\textwidth]{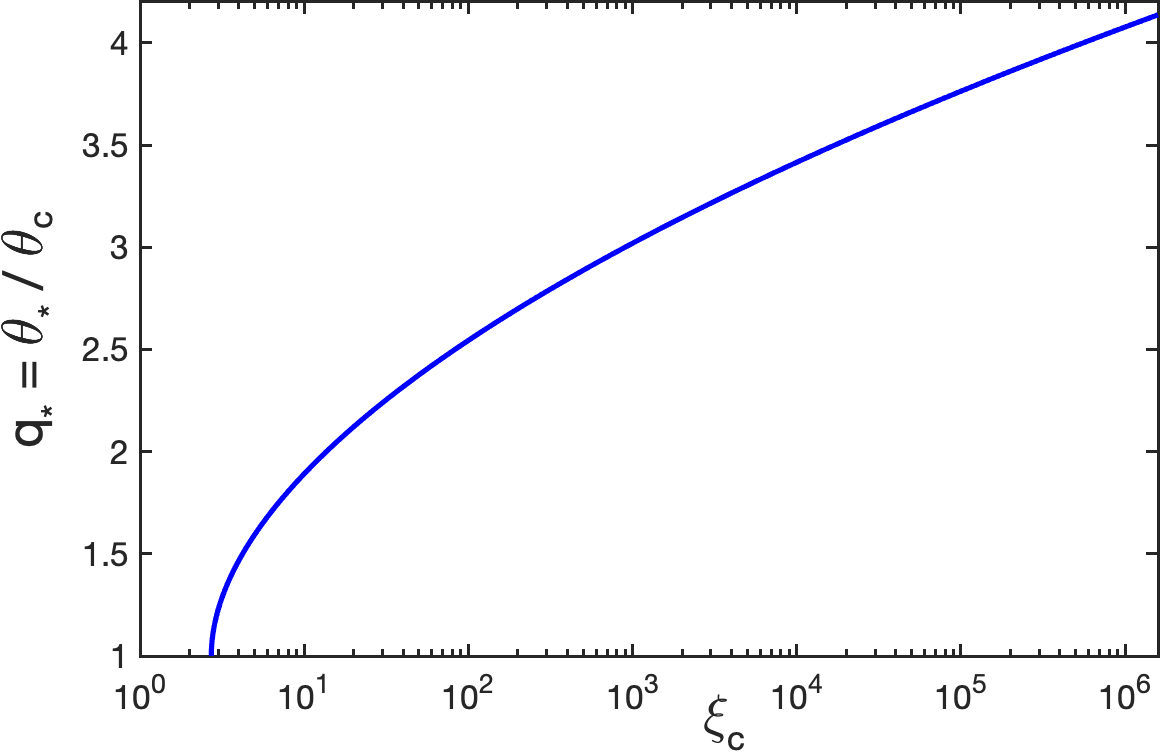}
		\vspace{0.22cm}\\
		\includegraphics[width=0.45\textwidth]{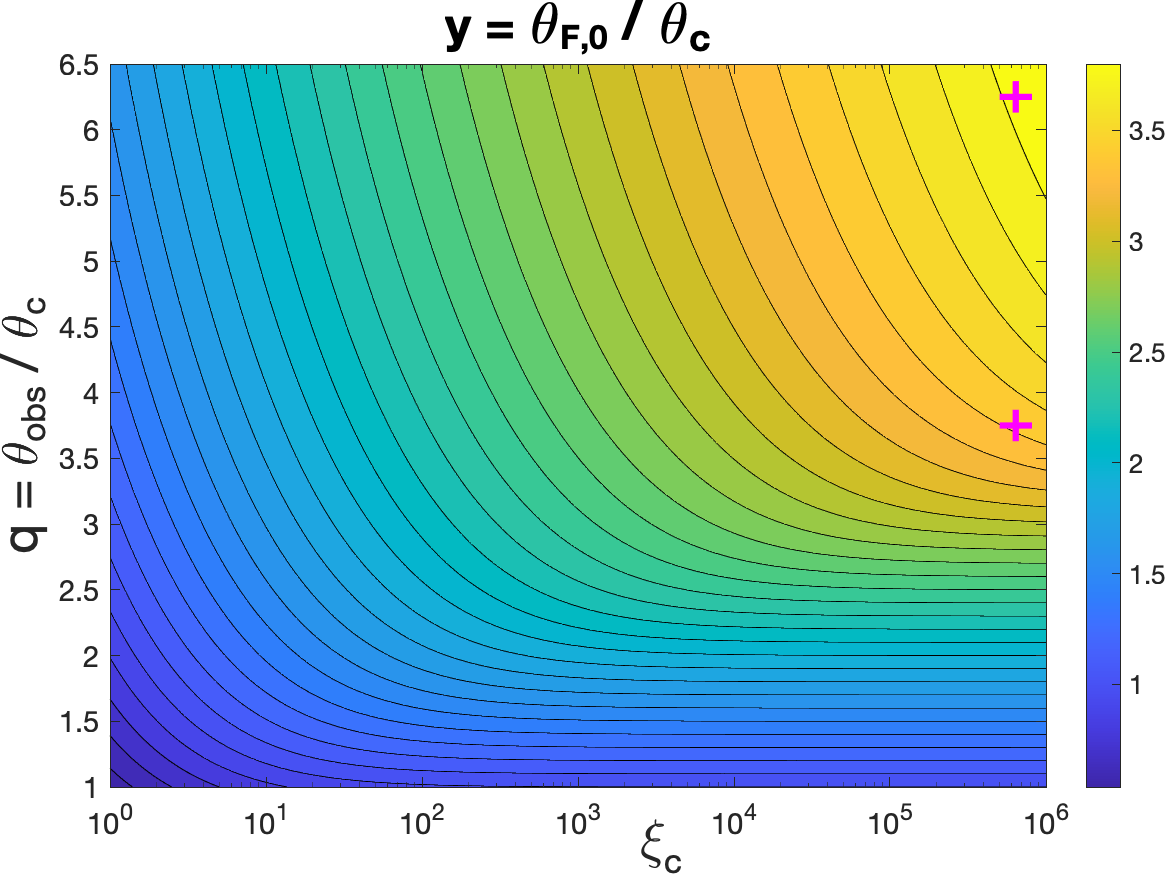}
		\vspace{0.22cm}\\
		\includegraphics[width=0.45\textwidth]{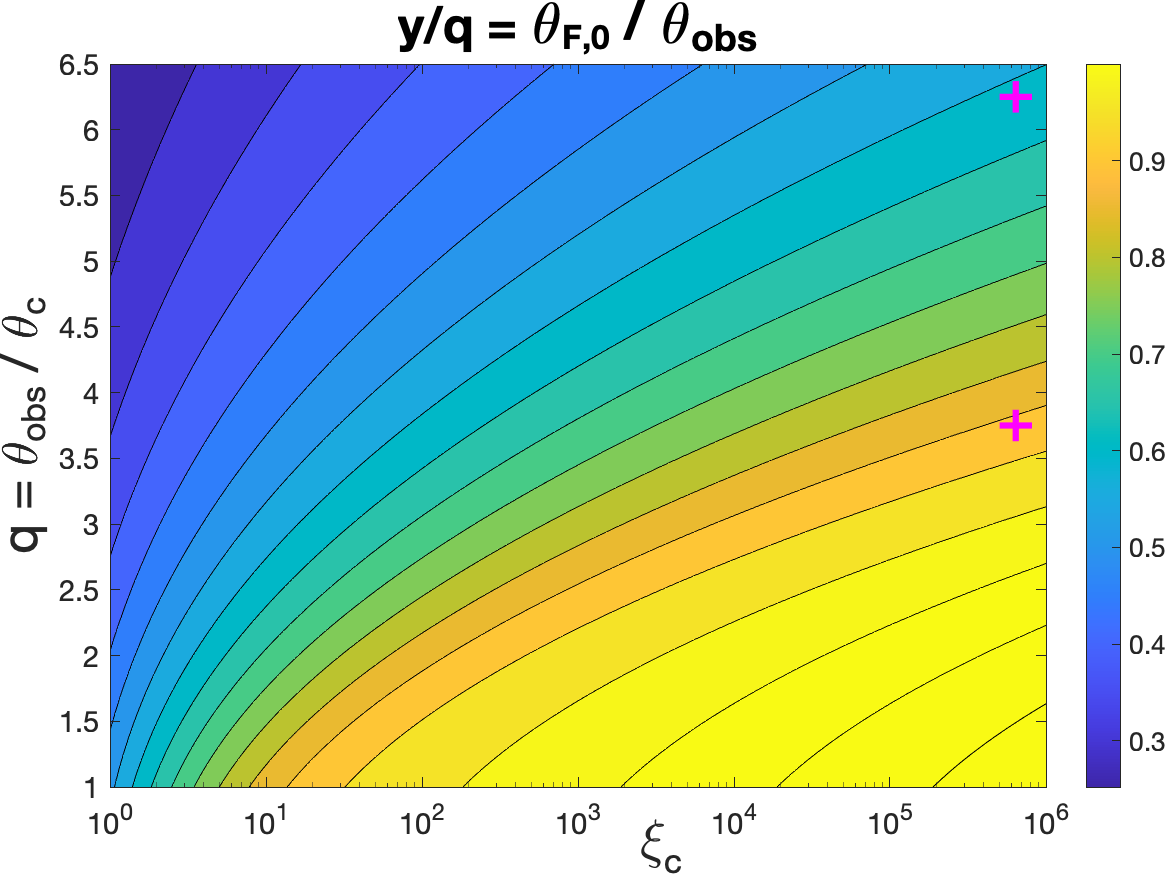}
		\vspace{-0.35cm}\\
		\caption{Some key parameters for a Gaussian jet.
			\textbfit{Top}: The angle $\theta_*$ defined by $\theta_*\Gamma_0(\theta_*)=1$ (a solution existes for $\xi_{\rm c}\geq e\approx2.718$) normalized by $\theta_{\rm c}$ as a function of $\xi_{\rm c}=(\Gamma_{\rm c,0}\theta_{\rm c})^2$ (see Eq.~(\ref{eq:Gaussian_qs})). \textbfit{Middle}: contour map of the normalized angle $y=\theta_{F,0}/\theta_{\rm c}$ in the $\xi_{\rm c}$-$q$ plane, for PLS G and $p=2.2$ (see Eq.~(\ref{eq:thetaFprox_Gaussian})); contours are at intervals of 0.1 from $y=0.5$ to $y=3.8$ while the two magenta plus symbols in the last two panes are for the two lightcurves shown in Fig.~\ref{fig:Gausslc}.
			\textbfit{Bottom}: a similar contour map of $y/q=\theta_{F,0}/\theta_{\rm obs}$ with contours at $y/q=0.25$:0.05:0.95,\,0.99,\,0.999,\,0.9999,\,0.99999.}  
		\label{fig:Gaussian_qs_y}
	\end{figure}

	In the ultra-relativistic limit, the equation for $\theta_{F,0}$ is given by
	\begin{equation}
	\label{eq:thetaFprox_Gaussian}
	\left(1-\frac{\lambda_{\epsilon}}{\lambda_{\mathcal{D}}}\right)(q\!-\!y)^2+2\frac{q\!-\!y}{y}=\left(1+\frac{\lambda_{\epsilon}}{\lambda_{\mathcal{D}}}\right)\frac{e^{y^2}}{\xi_{\rm c}}\quad,\quad y\equiv\frac{\theta_{F,0}}{\theta_{\rm c}}
	\end{equation}
	Since closed form algebraic solutions are not available for $\theta_*, \theta_{F,0}$ in the Gaussian structure case, we plot the numerical solutions for those parameters in Fig. \ref{fig:Gaussian_qs_y}.
	
	\section{Inferences from observations}
	\label{sec:Inferences}
	The relationship between the observed characteristic times and fluxes obtained for the different lightcurves discussed in \S \ref{sec:Modelling} can be used to infer some of the defining physical properties of a GRB.
	
	\subsection{The temporal slopes}
	\label{sec:slopes}
	In all the cases considered here, we have found a shallow portion of the lightcurve, with $F_\nu\propto t^{\alpha}$, that is governed by the temporal evolution of $\theta_{\rm min}(t)$.
	The derivation of $\alpha$ for PLS G is given in GG18. Applying the same derivation we provide here the values also for the other synchrotron PLS in Table~\ref{tbl:alpha}. In general, relating $\alpha$ to $\beta$ can provide a closure relation between the temporal and spectral slopes of the type $\alpha(\beta,a,k)$ which may be used to test the validity of the model to observations (see e.g. \citealt{Racusin2009,Ryan2019}).

	In addition, all cases exhibit an early phase, with $F_\nu\propto t^{\alpha_r}$, that corresponds to material that has not yet been decelerated and a late decline with $F_{\nu}\propto t^{\alpha_f}$.  
	The value of $k,p$ can be inferred from $\alpha_r,\alpha_f$. For example, for PLS G and $k=0$, assuming no lateral-expansion of the jet after the jet-break and as long as the flow is still relativistic, $\alpha_f=-3p/4$ \footnote{Immediately after the peak, the lightcurve is slightly steeper due to ``limb-brightening'' effect, \citep{Granot07}} and $\alpha_r=3$ (values for general values of $k$ and other synchrotron PLS are given in Table~\ref{tbl:alpha}). The value of $p$ can also be extremely well constrained from the spectrum, which is independent of the assumption on lateral expansion. In fact the observations of GRB 170817A spanning all the way from the radio to the X-ray band, and revealing a spectrum consistent with a single power-law segment in that frequency range, provided an unprecedented accuracy in determining $p=2.17$ \citep[e.g.,][]{Margutti2018,DAvanzo2018,Troja2018,Lyman2018,Resmi2018}.
	
	In principle the value of $a$ can be inferred directly from the asymptotic temporal index $\alpha$ during the rise to the peak that is governed by the angular effect \citep{Gill-Granot-18}, e.g. for PLS G (and $k=0$ corresponding to a uniform medium, as relevant for short GRBs),
	\begin{equation}
	a = \frac{8(3-k)}{4(3-k)-4\alpha-k(p+1)}\;\xrightarrow{k\rightarrow0}\;
	\frac{6}{3-\alpha}\ .
	\end{equation}
	However, in practice the dynamical range is limited and we are rarely deep in the limit $\theta_{\rm obs}\gg \theta_{\rm min}\gg \theta_{\rm c}$ for which this analytic result holds. For example, in GRB$\,$170817A/GW$\,$170817 the above analytic expression gave $a\approx2.7$ while a direct fit to the lightcurve gave $a\approx4.5$ \citep{Gill-Granot-18} since a steeper angular profile is required in order to compensate for the limited dynamical range.
	
	\begin{table*}
		\caption{Values of the temporal slopes for the different synchrotron PLS.}
		\centering
		\resizebox{0.5\textwidth}{!}{
			\begin{threeparttable}	\begin{tabular}{cccccc}\hline	
					PLS &  $\beta$ & $\alpha_i$ & $\alpha_d$ & $\alpha$  & $\alpha_f$\\ \hline 
					D & $-\frac{1}{3}$ & $3-k/2$ & $\frac{2-k}{4-k}$ &  $\frac{8(k-3)-a(16k/3-12)}{4a}$ & $\frac{1}{k-4}$ \vspace{0.2cm}\\ 
					E & $-\frac{1}{3}$ &  $11/3-2k$ & $\frac{2-3k}{3(4-k)}$ & $\frac{8(k-3)-a(8k-44/3)}{4a}$ & $\frac{7}{3(k-4)}$ \vspace{0.2cm} \\
					F & $\frac{1}{2}$ &  $2-3k/4$ & $-1/4$ & $\frac{8(k-3)-a(3k-8)}{4a}$ & $\frac{16-5k}{4(k-4)}$ \vspace{0.2cm} \\
					G & $\frac{p-1}{2}$ &  $3-k(p+5)/4$ & $\frac{k(3p-5)-12(p-1)}{4(4-k)}$ & $\frac{8(k-3)-a((5+p)k-12)}{4a}$ & $\frac{k+12p-3kp}{4(k-4)}$ \vspace{0.2cm} \\
					H & $\frac{p}{2}$ &   $2-k(p+2)/4$ & $(2-3p)/4$ & $\frac{8(k-3)-a((2+p)k-8)}{4a}$ & $\frac{4+12p-k(2+3p)}{4(k-4)}$  \\
					\hline
					\label{tbl:alpha}
				\end{tabular}
			\end{threeparttable}
		}
	\end{table*} 
	
	In cases 1A, 2, 3, there is another declining phase after the first peak. The asymptotic slope of this decline is similar to that of an on-axis afterglow and depends on the observed synchrotron PLS. We shall denote it by $\alpha_d$ where $F_{\nu}\propto t^{\alpha_d}$. As an example, for PLS G and $k=0$, $\alpha_d=-3(p-1)/4$ (see Table~\ref{tbl:alpha} for other cases). We caution the reader that in practice, if the dynamical range between the first peak and the dip is not sufficiently large, the observed decline may be significantly flatter.
	
	\subsection{timescales}
	All the critical timescales in determining the observed afterglow lightcurve are proportional to $t_{\rm dec,c}\propto (\epsilon_{\rm c}/A)^{1\over 3-k}\Gamma_{\rm c,0}^{2k-8 \over 3-k}$. In particular, there is a degeneracy between $\epsilon_{\rm c}, A$ which makes it challenging to infer any one of these parameters on its own. However, since they appear in all the timescales through the same scaling, comparing the {\it ratio} of different lightcurve characteristic times is particularly useful for robustly inferring physical properties. We discuss these inferences below.
	
	The time of the main peak in the lightcurve $t_{\rm{pk}}$ for off-axis observers ($q>1$) can be identified with time at which the jet's core becomes visible, i.e. $\theta_{F}(t_{\rm{pk}})=0$. Assuming that the core is initially not visible ($\theta_{F,0}>\theta_{\rm c}$), this time is approximately the same as when $\theta_{\rm min}\to 0$ using the broken power-law description of $\Gamma(t)$
	\begin{equation}\label{eq:t_pk}
	\tilde{t}_{\rm{pk}} = 2^{-a/[2(3-k)]}\xi_{\rm c}^{4-k \over 3-k }q^{2(4-k)/(3-k)}\ .
	\end{equation}
	
	In cases 1A, 2, 3 the lightcurve is double-peaked. In these cases, two other critical timescales appear, the time of the first peak, and the time of the dip between the two peaks. These can be estimated in the following way
	\begin{equation}
	\label{eq:t1pk}
	\tilde{t}_{\rm 1pk}=\tilde{t}_{\rm dec}(\theta_{F,0})\approx \tilde{t}_{\rm dec}(\theta_{\rm obs})= \left(1+q^2\right)^{[2(4-k)b-a]/2(3-k)}
	\end{equation}
	\begin{equation}
	\label{eq:tdip}
	\tilde{t}_{\rm dip}\approx \tilde{t}_{\rm beam}(\theta_{\rm obs})=2^{-a\over 2(3-k)}(1+q^2)^{-a\over 2(3-k)} q^{2(4-k)\over 3-k}\xi_{\rm c}^{4-k \over 3-k}
	\end{equation}
	where in the r.h.s. of equation \ref{eq:tdip}, we have used a slightly more accurate definition of $t_{\rm dip}$, namely that the extrapolations of the flux from the first and second peak match, i.e. $F_{\rm 1p}(t_{\rm dip}/t_{\rm 1pk})^{\alpha_d}=F_{\rm pk}(t_{\rm dip}/t_{\rm pk})^{\alpha}$ (the expressions for those fluxes are given in the next sub-section). We note that the difference between $\tilde{t}_{\rm beam}(\theta_{\rm obs})$ and the more accurate prescription are rather small, up to tens of percent throughout the majority of the parameter space.

	Assuming $k,a$ can be determined from the temporal slopes (see above) and at the limit $q\gg 1$ we find
	\begin{equation}
	q=\frac{\theta_{\rm obs}}{\theta_{\rm c}}\approx \bigg(\frac{t_{\rm pk}}{t_{\rm dip}}\bigg)^{3-k\over a}
	\end{equation}
	and
	\begin{equation}
	\xi_{\rm c}\approx2^{{a \over 2(4-k)}} \bigg(\frac{t_{\rm pk}}{t_{\rm 1pk}}\bigg)^{3-k\over 4-k}q^{2(b-1)-\frac{a}{4-k}}
	\end{equation}
	
	In case 1B, the time of the peak remains the same as in equation \ref{eq:t_pk}, but the times of the first peak and the dip are no longer relevant. Instead, a new timescale appears, which is the time at which the shallow rise ($F_{\nu}\propto t^{\alpha}$) starts, $t_{\rm dec}(\theta_{F,0})$. In this regime, $\theta_{F,0}$ can be well estimated by Eq.~(\ref{eq:thetaFprox}). A slightly less accurate but easier approximation for the time of the initial rise is given by $t_{\rm dec}(\theta_{\rm min,0})$. When $\theta_*<\theta_{\rm obs}$, the latter is roughly given by
	\begin{eqnarray}
	\label{eq:thetamin0approx}
	&  \theta_{\rm min,0}\approx \theta_{\rm c} \xi_{\rm c}^{\frac{1}{2b}}\left(q-\xi_{\rm c}^{\frac{1}{2b}}q^{\frac{1}{b}}\right)^{\frac{1}{b}} \Longrightarrow \\&  t_{\rm dec}(\theta_{\rm min,0})=\left[1+\xi_{\rm c}^{\frac{1}{b}}\left(q-\xi_{\rm c}^{\frac{1}{2b}}q^{\frac{1}{b}}\right)^{\frac{2}{b}}\right]^{2(4-k)b-a \over 2(3-k)}.
	\end{eqnarray}
	Finally, the most straight-forward but least accurate expression for the initial rise is given by
	$\tilde{t}_{\rm dec}(\theta_{\rm min,0})\approx \tilde{t}_*(\theta_{\rm obs}/\theta_*)^{[2(4-k)b-a]/(3-k)b}$ (see \S \ref{sec:Modelling}) or equivalently
	\begin{equation}
	\tilde{t}_{\rm dec}(\theta_{\rm min,0})\approx (\xi_{\rm c}q^2)^{[2(4-k)b-a]/[2b(3-k)]}
	\end{equation}
	In this case we obtain the following relation between the observables and the physical parameters
	\begin{equation}
	\label{eq:xic}
	\xi_{\rm c}q^2=\bigg(\frac{t_{\rm pk}}{t_{\rm dec}(\theta_{\rm min,0})}\bigg)^{2b(3-k)\over a} 2^{b}
	\end{equation}
	
	\subsection{Fluxes}
	In cases 1A, 2, 3, there are three characteristic fluxes. For concreteness we assume that the observed band is in PLS G. The flux at $t_{\rm pk}$ has been well studied in the literature (e.g. \citealt{Nakar2002}). For $k=0$ the result is 
	\begin{equation}
	\label{eq:Fpk}
	F_{\rm pk}\propto \epsilon_{\rm c} \theta_{\rm c}^2 n^{p+1\over 4}\epsilon_e^{p-1}\epsilon_B^{p+1\over 4}\nu^{1-p\over 2}d_L^{-2}\theta_{\rm obs}^{-2p}
	\end{equation}
	where $n$ is the particle number density of the circumburst medium, $d_L$ is the luminosity distance of the GRB, and $\epsilon_e$ and $\epsilon_B$ are the 
	shock microphysical parameters representing the fractions of the total internal energy density behind the shock deposited in relativistic electrons and magnetic fields respectively. 
	Since the peak flux is degenerate between several of the bursts' properties, we consider, as for the timescales, the fluxes relative to the peak flux.
	The flux at the first peak can be approximated by noticing that the time and flux of the later peak are directly related to the time and flux at the moment of the jet break (i.e. when $\Gamma(\theta\!=\!0,t_{\rm j,b})\theta_{\rm c}=1$). Assuming no lateral expansion, the appropriate expressions are $t_{\rm pk}/t_{\rm j,b}\approx (\theta_{\rm obs}/\theta_{\rm c})^{8/3}$, $F_{\rm pk}/F_{\rm j,b}\approx (\theta_{\rm obs}/\theta_{\rm c})^{-2p}$ \citep{Nakar2002}. The time and flux at $t_{\rm j,b}$ can then be extrapolated back to $t_{\rm 1pk}$, using the standard spherically symmetric pre-deceleration description of the flux (e.g. \citealt{GS02}) and using $E_{\rm iso}\sim 4\pi \epsilon(\theta)$. The result is
	\begin{equation}
	\label{eq:F1pk}
	F_{\rm 1 pk}\approx F_{\rm pk} \bigg(\frac{\theta_{\rm obs}}{\theta_{\rm c}}\bigg)^{8-a(3+p)\over 4}\bigg(\frac{t_{\rm 1pk}}{t_{\rm pk}}\bigg)^{3(1-p)\over 4}\approx
	2^{-{a\over 8}}\xi_{\rm c} q^{4-a-2b}
	\end{equation}
	where in the r.h.s. we have plugged the asymptotic expressions for the timescales and $p\approx 2$ for clarity. This demonstrates that it is possible for the first peak to be brighter than the second one for large enough $q$ and for sufficiently small $a,b$.
	Out of the five physical quantities $F_{\rm pk},F_{\rm 1pk},t_{\rm pk},t_{\rm dip},t_{\rm 1pk}$ only four are truly independent (see discussion in the previous sub-section regarding $t_{\rm dip}$). There is therefore a choice of which quantities to use, depending on how well they can be determined and what physical quantity is attempted at being deduced.
	
	The flux at $t_{\rm dip}$ is given by a direct extrapolation from $t_{\rm pk}$,
	\begin{equation}
	F_{\rm dip}=F_{\rm pk} \bigg(\frac{t_{\rm dip}}{t_{\rm pk}}\bigg)^{\alpha}
	\end{equation}
	Clearly, $F_{\rm dip}$ does not provide independent information to that obtained from $t_{\rm dip}/t_{\rm pk}$ and the observed temporal slope. The ratio $F_{\rm 1pk}/F_{\rm pk}$ is however more illuminating and provides an independent estimate of the observation angle
	\begin{equation}
	q=\bigg(\frac{F_{\rm 1pk}}{F_{\rm pk}}\bigg)^{4\over 8-a(3+p)}\bigg(\frac{t_{\rm pk}}{t_{\rm 1pk}}\bigg)^{3(1-p)\over 8-a(3+p)}
	\end{equation}
	
	In case 1B, one can express the flux at $t_{\rm dec}(\theta_{F,0})$ as
	\begin{equation}
	F(t_{\rm dec}(\theta_{F,0}))=F_{\rm pk} \bigg(\frac{t_{\rm dec}(\theta_{F,0})}{t_{\rm pk}}\bigg)^{\alpha}.
	\end{equation}
	As in the case of $F_{\rm dip}$, this does not provide additional information to that given by the equations for the temporal slope and for $t_{\rm dec}(\theta_{F,0})$.
	
	\section{GRB 170817A as a test case}
	\label{sec:170817}
	We have found in this work four lightcurve regimes with two main qualitative types of GRB afterglows, with and without a double peak. The qualitative difference between these regimes has some straight-forward implications on the physical parameters.
	
\begin{figure}
		\centering
		\includegraphics[width=0.4\textwidth]{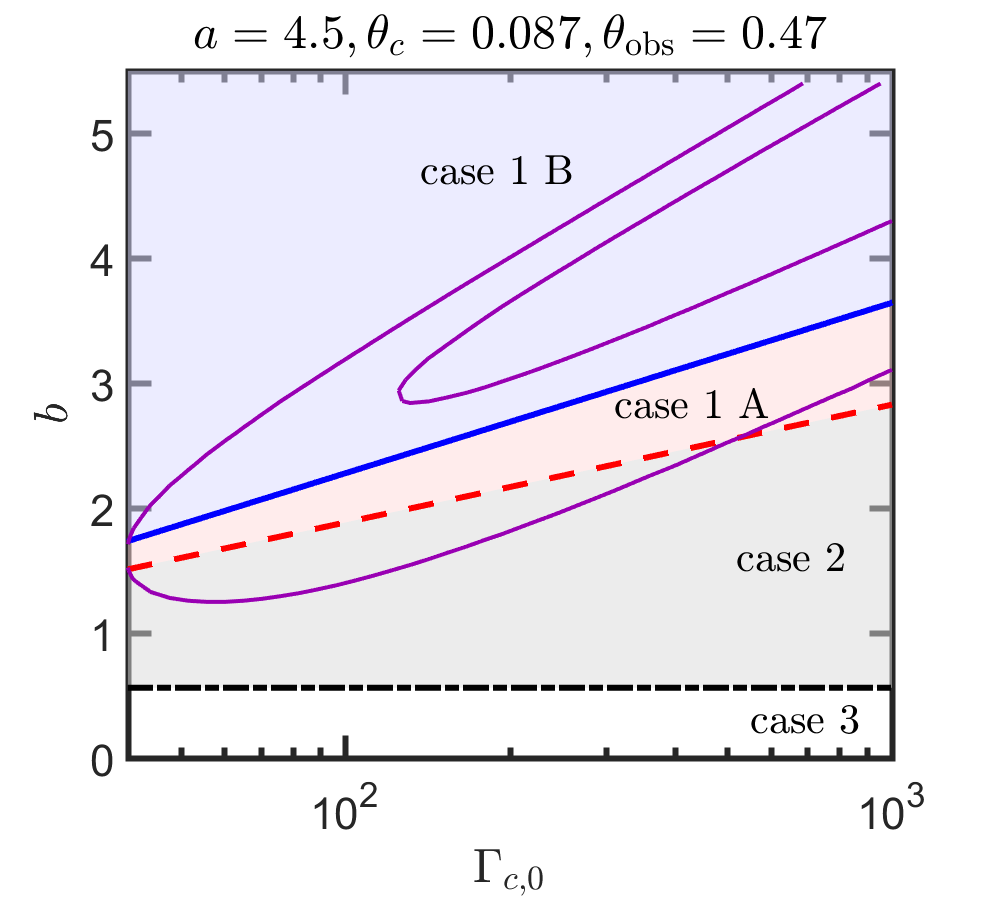}
		\caption{
			Allowed parameter space for the different types of lightcurves presented in this paper. The blue solid line depicts $\theta_{\rm obs}=\theta_*$, the red dashed line depicts $b=b_c$ and the dot dashed black line depicts $b=b_a$. Results are plotted with $\theta_{\rm c}=0.087,\theta_{\rm obs}=0.47$ as inferred for GRB$\,$170817A (as well as $a=4.5$ for the purpose of distinguishing between case 2 and case 3, the results depend very weakly on the specific value). lightcurves with a single peak (case 1B) require large $b$ and / or small $\Gamma_{\rm c}$ (above the blue line). As the distance from the blue line increases the early peak emerges and becomes gradually stronger. Overplotted in purple are the same $\chi^2_\nu \leq 3.2$ and $\chi^2_\nu \leq 2.7$ model fitting contours shown in Fig. \ref{fig:paramGW170817-fit}.}
		\label{fig:paramspace}
	\end{figure}

	Consider for example, a situation in which one excludes with confidence the existence of a double peak in a given GRB afterglow. Indeed, this may be the case for GRB 170817A in which the first detections occurred at $\sim$10 days after the burst, while early observations yielded a strong upper limit on the flux starting from $\sim 1$ day after the trigger. What can be learned from this observation? In order to avoid a first peak, the conditions must be close to those corresponding to case 1b. Namely (i) $b>b_c(\Gamma_{\rm c,0},\theta_{\rm c})$ and (ii) $\theta_{\rm obs}>\theta_*(\Gamma_{\rm c,0},\theta_{\rm c})$. Since the values of $\theta_{\rm c}\approx 0.087, \theta_{\rm obs}\approx 0.47$ are relatively well constrained from the combination of the superluminal motion observation and the time of the observed peak \citep{Mooley2018,Pooley2018,Gill+19} and since $a$ can be reasonably well constrained from the shallow rise of the lightcurve towards the peak (see \S \ref{sec:slopes}), it is useful to describe the parameter space corresponding to the different lightcurve regimes in terms of $b,\Gamma_{\rm c,0}$. The results are shown in Fig. \ref{fig:paramspace}. Large $b$ and / or small $\Gamma_{\rm c,0}$ are needed to completely avoid the first peak. In general, as $b$ becomes smaller, and / or $\Gamma_{\rm c,0}$ becomes larger a second peak emerges and gradually becomes stronger.

\begin{figure}
		\centering
		\includegraphics[width=0.48\textwidth]{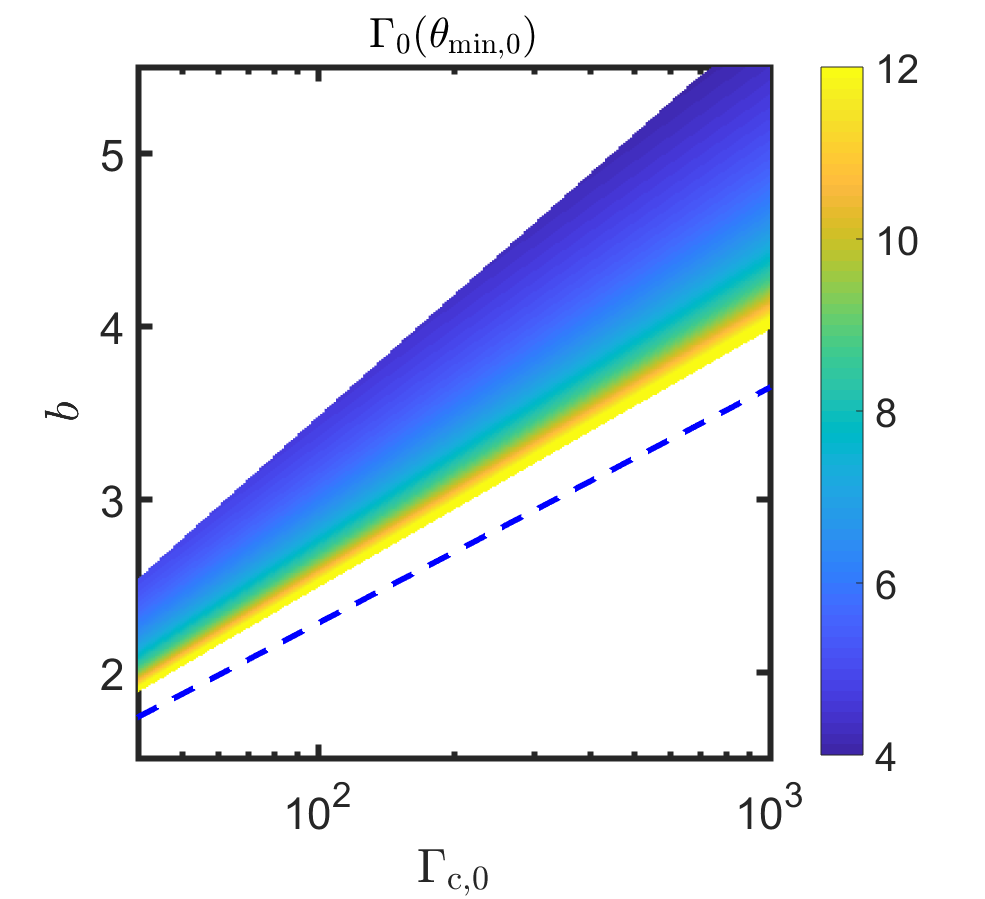}\\
		\caption{Allowed parameter space provided by the requirements that only one peak is seen in the lightcurve with $t_{\rm pk}>7 t_{\rm dec}(\theta_{\rm min,0})$ and assuming  $a=4.5, \theta_{\rm c}=0.087,\theta_{\rm obs}=0.47$. Colour represents the value of $\Gamma_0(\theta_{\rm min,0})$ which in this regime is approximately the initial Lorentz factor of the material dominating the early lightcurve. A dashed blue line depicts the boundary between case 1B and case 1A (see Fig. \ref{fig:paramspace}).
		}
		\label{fig:paramGW170817}
	\end{figure}

	In GRB 170817A, the peak of the lightcurve occurred at $t_{\rm pk}\approx 150$ days, while the beginning of the shallow rise started at $\approx 10-20$ days. The large span of time between the beginning of the shallow rise and the eventual peak, provides another constraint on the allowed parameter space. Assuming no lateral expansion and conservatively taking $t_{\rm pk}/t_{\rm dec}(\theta_{\rm min,0})>7$, we may use Eq.~(\ref{eq:xic}) to further constrain the allowed parameter space. The results are shown in Fig. \ref{fig:paramGW170817}. 
	
\begin{figure}
		\centering
		\includegraphics[width=0.48\textwidth, trim=0 0 58 0]{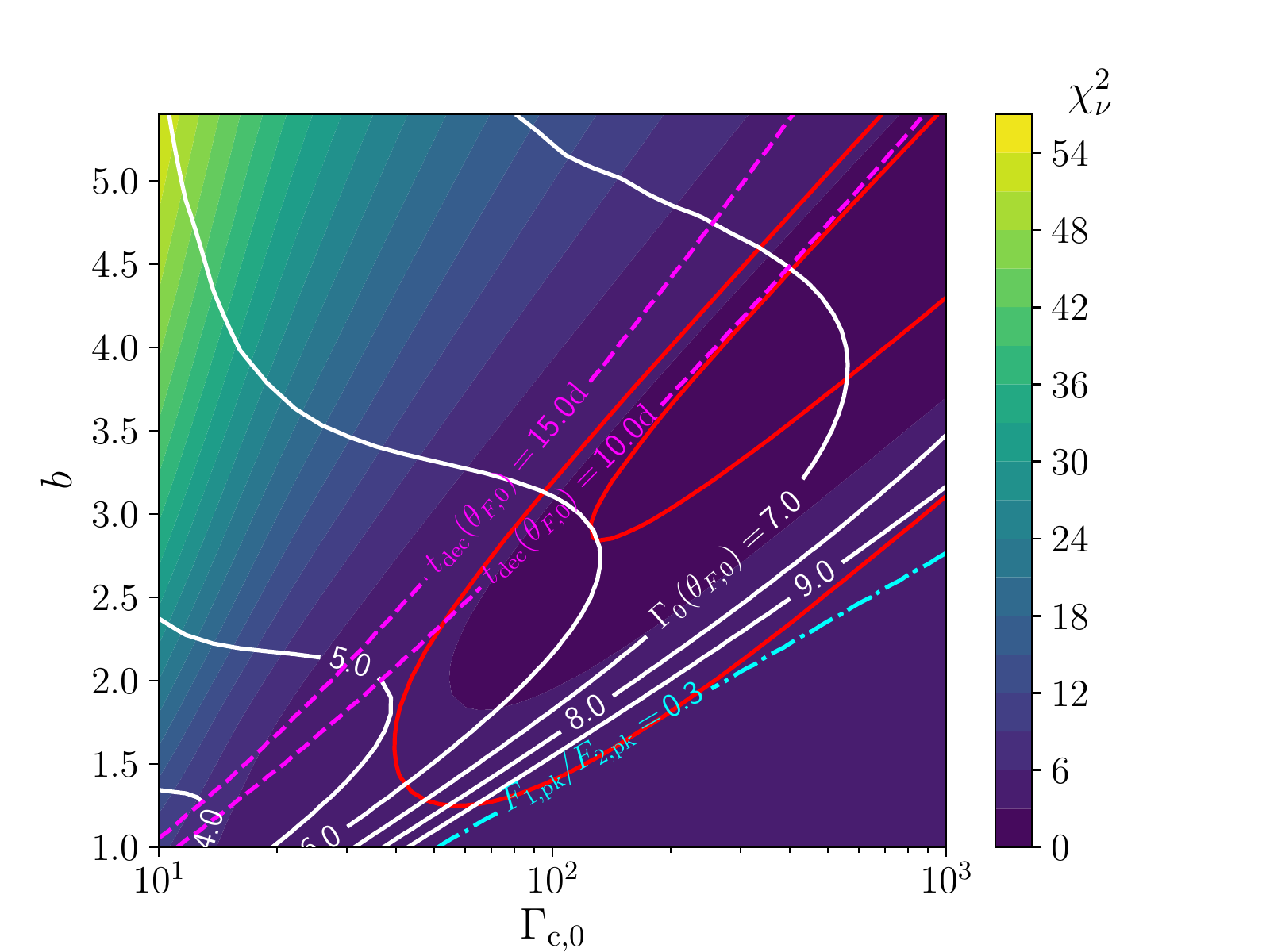} \\
		\includegraphics[width=0.48\textwidth]{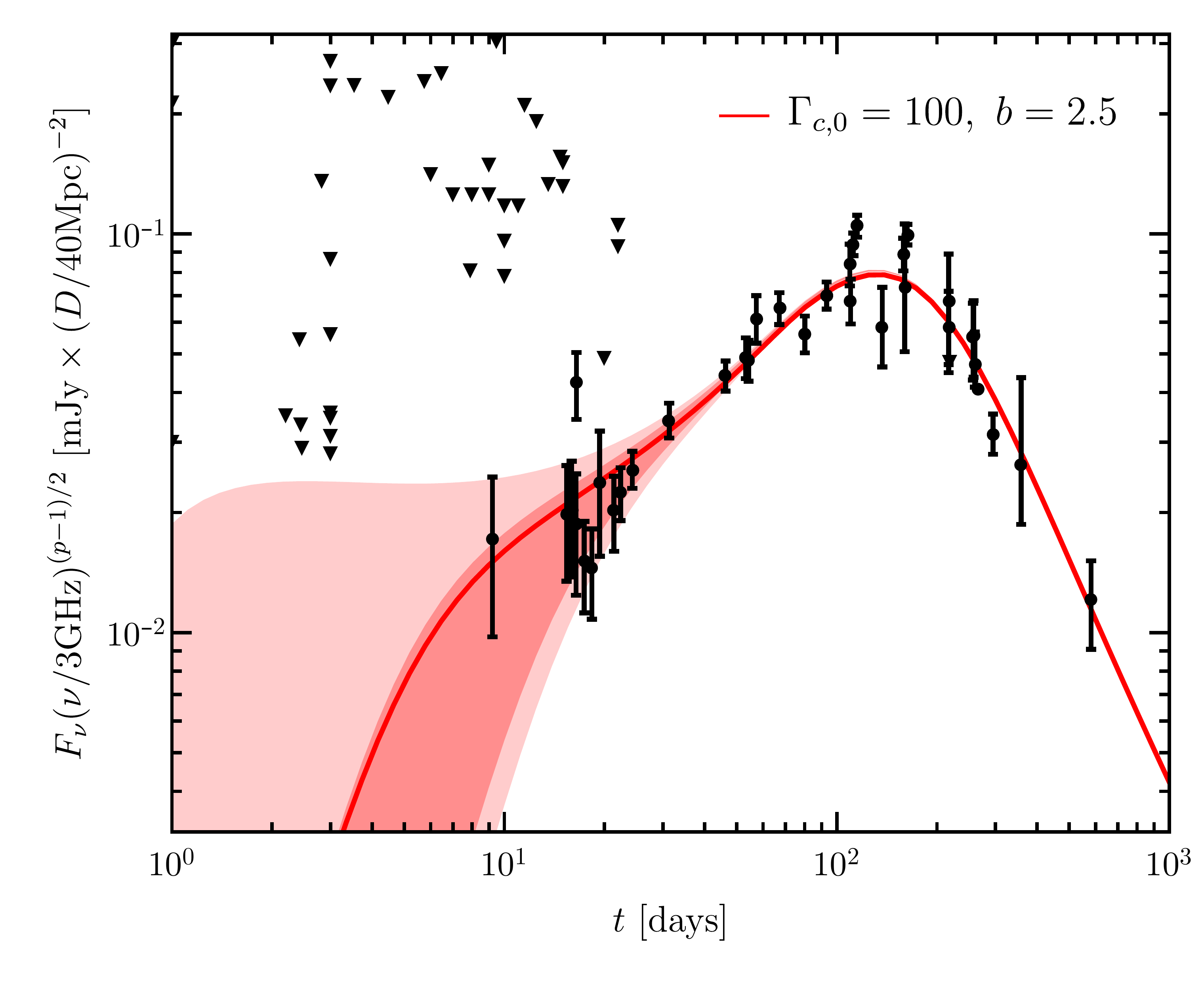}
		\caption{{\bf Top:} Reduced chi-square ($\chi^2_\nu$) contour map obtained by fitting the PL model of GG18 
			(with the same $a,\theta_{\rm c},\theta_{\rm obs}$ as in Fig.\ref{fig:paramspace}) 
			to the afterglow data of GRB$\,$170817A. The outer and inner solid red curves encompass 
			regions in the parameter space that give the best-fit solution with $\chi^2_\nu \leq 3.2$ and $\chi^2_\nu \leq 2.7$, respectively. 
			These regions are constrained from above by dashed magenta lines for which the parameter space below the lines always 
			yield the condition that $t_{\rm dec}(\theta_{F,0})<t_1$, where $t_1$ are representative times of the early afterglow 
			observations. Another constraint is shown by a dot-dashed cyan line for which the region above the line always yield 
			the flux ratio $F_{1,{\rm pk}}/F_{2,{\rm pk}}<0.3$. Solid white contour lines indicate the initial bulk Lorentz factor of the 
			initial angle that dominated the flux. 
			{\bf Bottom:} PL jet model lightcurve fit to the afterglow data of GRB$\,$170817A, with upper limits marked with downward triangles. 
			The solid red curve shows the best-fit solution of \citet{Gill-Granot-18,Gill+19}. The lighter and darker shaded red regions 
			encompass lightcurves obtained for $\{\Gamma_{\rm c,0}, b\}$ values with best-fit $\chi^2_\nu \leq 3.2$ and $\chi^2_\nu \leq 2.7$, 
			respectively.
		}
		\label{fig:paramGW170817-fit}
	\end{figure}

	In Fig. \ref{fig:paramGW170817-fit}, we show the reduced chi-square ($\chi^2_\nu$) contour map (top-panel) in the plane of $\{\Gamma_{\rm c,0}, b\}$ as obtained from a PL structured jet model fit of GG18, with $a=4.5, \theta_{\rm c}=0.087, \theta_{\rm obs}=0.47$, to the afterglow data of GRB$\,$170817A. In the bottom-panel, we show the best-fit lightcurve from GG18 and \citet{Gill+19} along with shaded regions that encompass lightcurves obtained for different values of $\{\Gamma_{\rm c,0}, b\}$ that correspond to $\chi^2_\nu\leq3.2$ and $\chi^2_\nu\leq2.7$. The parameter space providing the best fits for these models agrees well with the space given by the requirement of having one peak with $t_{\rm pk}>7t_{\rm dec}(\theta_{\rm min,0})$. Both calculations lead to a narrow allowed region in the  $\{\Gamma_{\rm c,0},b\}$ parameter space, which represents a constraint on $\Gamma_0(\theta_{\rm min,0})$. If the latter is too large then there will be an early peak that quickly becomes too bright compared to the available limits. Instead, if the Lorentz factor of the material dominating the early lightcurve is too small, then this material takes too long to decelerate and the shallow rise doesn't last for long enough. The value required by the conditions outlined above leads to $\Gamma_0(\theta_{\rm min,0})\approx 5-7$. From the $\chi^2_\nu$ map (regions within the red contours in the top panel) and the corresponding shaded red regions in the lightcurve  plot, 
	it is clear that PL structured jet models with $b\lesssim1.2$ and $\Gamma_{\rm c,0}\lesssim 40$ would not fit the afterglow data of GRB$\,$1701817A,  particularly 
	at early times at $t<40\,$days.
	
	The value of $\Gamma_0(\theta_{\rm min,0})$ can be independently understood from the observations of superluminal motion in GRB 170817A \citep{Mooley2018,Ghirlanda2019}. These observations have revealed that the flux centroid of GRB 170817A was moving with an apparent velocity of $\approx 4c$ around the time of the lightcurve peak. The implication is that $\Gamma(\theta_{\rm c},t_{\rm pk})\approx 4$. This result can be related to the Lorentz factor along the direction initially dominating the lightcurve, by using the definition of $\theta_{\rm min}$ (Eq.~(\ref{eq:thetaminfull})), $\Gamma_0(\theta_{\rm min,0})=f \Gamma(\theta_{\rm c},t_{\rm pk})$, where $f= (\theta_{\rm obs}-\theta_{\rm c})/(\theta_{\rm obs}-\theta_{\rm min,0})$. Since by definition $f\geq 1$, this immediately suggests that $\Gamma_0(\theta_{\rm min,0})\gtrsim 4$. In regime 1B, $\theta_{\rm min,0}\ll \theta_{\rm obs}$, implying $f\approx1$.
	Using Eq.~(\ref{eq:Gammathetat}) we can obtain a rough estimate for $f$ using the ratio $t_{\rm pk}/t_{\rm dec}(\theta_{\rm min,0})\approx (\theta_{\rm min,0}/\theta_{\rm c})^{a/(3-k)}$. The condition $t_{\rm pk}/t_{\rm dec}(\theta_{\rm min,0})>7$ then implies $(\theta_{\rm min,0}/\theta_{\rm c})\gtrsim 3.7$.
	In reality, as shown in Fig. \ref{fig:thetaminthetaF}, the decay of $\theta_{\rm min}(t)$ is less steep than the asymptotic PL decay. Taking, for example, $\theta_{\rm obs},\theta_{\rm c}, a$ as above, as well as $\Gamma_{\rm c,0}=300, b=4$ (where the condition $t_{\rm pk}/t_{\rm dec}(\theta_{\rm min,0})>7$ is satisfied, as shown in Fig. \ref{fig:paramGW170817}) we find $(\theta_{\rm min,0}/\theta_{\rm c})\approx 2.8$.
	Plugging this back into $f$ (and using our values for $\theta_{\rm obs}/\theta_{\rm c}$) we find $f\approx 1.7$. Overall, we conclude that $\Gamma_0(\theta_{\rm min,0})\approx 7$, which is consistent with what we have found above from the more detailed calculation.
	Finally, the value of $\Gamma_0(\theta_{\rm min,0})$ is also consistent with the limits for the material dominating the prompt material which from compactness arguments leads to $\Gamma\gtrsim 2-3$ \citep[e.g.,][]{Kasliwal2017,Matsumoto2019}.
	Summarizing, we find three independent constraints on $\Gamma_0(\theta_{\rm min,0})$ (from the lightcurve analysis, from superluminal motion and from compactness limits) which are all in broad agreement with each other. This lends credence to the angular jet structure models considered in this work. Furthermore, it outlines easily applicable consistency checks that can be used for future events to compare between the different models.
	
	\section{Conclusions}
	\label{sec:conc}
	We studied analytically the shapes of GRB afterglows that arise from structured jets viewed off-axis. We found qualitatively different types of lightcurves that may be viewed, depending on the jet properties and on the line of sight to the observer. Most notably, the lightcurve may be either singly or doubly peaked, depending on whether $\theta_{\rm obs}$ is (correspondingly) larger or smaller than a critical angle, $\theta_*$ \footnote{When $\theta_*$ becomes ill-defined or $\theta_*\gtrsim 1$ the lightcurve is always doubly peaked, see \S \ref{sec:Modelling}}.
	
	GRB afterglow fitting involves many unknown model parameters that quantify the jet properties, the surrounding medium, the shock microphysics, and the observer's viewing angle. Many of these different properties or model parameters are degenerate and cannot be uniquely determined or constrained, even with a very good set of observations, such as e.g. in GRB\,170817A\,/\,GW\,170817. However, focusing on the {\it shape} of the lightcurves, rather than the absolute normalizations of the flux and\,/\,or timescales, immediately removes the dependence on many of those parameters, and can provide very significant constraints on a sub-set of them.
	
	In the single peak scenario (case 1B) the shape of the lightcurve provides four scale-free observables: the early rise slope, $\alpha_r$, the shallow rise slope, $\alpha$, the final decline slope, $\alpha_f$ and the ratio between the start and end of the shallow rise phase, $t_{\rm dec}(\theta_{\rm min,0})/t_{\rm pk}$.
	The first three can constrain $k,p,a$ \footnote{In reality it may be challenging to observe the early rise slope. In X-rays it may be overshadowed by `internal' emission (i.e. from below the forward shock) associated with central engine activity \citep{Lu2015,BM2017}. In the optical, it may be overshadowed by the kilonova emission \citep{Kasen2017}. Finally, in the radio it may be overshadowed by reverse shock emission \citep{LK2019}.}. The fourth condition then provides a specific relation between $\xi_{\rm c}, b, q$. If one has some additional knowledge of $\theta_{\rm c},\theta_{\rm obs}$, for example from superluminal motion observations and\,/\,or from the width of the lightcurve peak, then the $t_{\rm dec}(\theta_{\rm min,0})/t_{\rm pk}$ constraint can be reduced to a simple relation between $b$ and $\Gamma_{\rm c,0}$. This leads to an estimate of the initial Lorentz factor along the line of sight that can be independently tested by superluminal motion observations and\,/\,or compactness of the prompt emission. Applying this analysis to GRB 170817A we find the Lorentz factor of material moving along the line of sight to the observer to be $\Gamma_0(\theta_{\rm min,0})=5-7$ as well as $b\gtrsim 1.2, \Gamma_{\rm c,0}\gtrsim 40$.
	These are the first direct limits from a single event viewed off-axis, which show that the core must have been ultra-relativistic, with a much larger Lorentz factor than the material that dominated the observed emission. This has implications, for example, for the prompt emission phase of GRBs, in which, depending on the radiation mechanism, the Lorentz factor may have a profound effect on the $\gamma$-ray spectrum and lightcurve. For example, \cite{BN2019} have recently studied structured jet models and have shown that, at least in long GRBs, the efficiency of $\gamma$-ray production must be significantly diminished for $\Gamma_0(\theta_{\rm min,0})\lesssim 50$ (see also \citealt{Hascoet2014,Ghirlanda2018,Matsumoto2019B}), effectively shutting off the prompt emission far beyond the jet's core.
	
	In the double peak scenario the shape of the lightcurve depends on five independent scale-free observables. These can be chosen in different ways. One useful such set is the four temporal slopes: $\alpha_r,\alpha, \alpha_d,\alpha_f$ and the ratio of the the early and late peak times $t_{\rm 1pk}/t_{\rm pk}$. Once more, the slopes constrain $k,p,a$, while the time ratio provides a relation between $\xi_{\rm c}, b, q$. This relation can be reduced to a relation between $b$ and $\Gamma_{\rm c,0}$ in a similar way to that described above for the singly peaked scenario. It is worth noting that reverse shock emission may also lead to an early peak in the afterglow lightcurve, predominantly in the radio band \citep{Lamb2019}, which is physically distinct from the origin of the early peak described in this work. In case an early peak is detected, multi-wavelength observations and\,/\,or spectral analysis of the lightcurve, could potentially be used to distinguish between the different scenarios.
	
	Only a fraction of GW detected binary neutron star mergers will have a detectable electromagnetic signal \citep[e.g.][]{lamb2017,BPBG2019,Duque2019,Kathirgamaraju2019}. Current predictions suggest that between the main electromagnetic counterparts: the prompt GRB emission, the kilonova (and / or its afterglow) and the GRB afterglow, it is the latter that is likely to be detected in GW triggered events most often \citep{Duque2019}. It is therefore of great importance to understand what physical parameters can be directly probed by such detections. Indeed, the detection fraction alone, can be used to statistically constrain the energy of the explosions and the typical densities of the surrounding medium \citep[e.g.][]{duque2020,BP2019}. The analysis described in this work can be used to significantly enhance numerical fitting attempts by potentially eliminating large portions of the initial parameter space as well as aiding with providing a physical interpretation and testable predictions for the model.
	
	\vspace{0.35cm}
	\section*{Acknowledgements}
	PB's research was funded in part by the Gordon and Betty Moore Foundation through Grant GBMF5076. This research was also supported by the ISF-NSFC joint research program (grant No. 3296/19; RG, JG). PB thanks Wenbin Lu, Geoffrey Ryan, Pawan Kumar and Ilaria Caiazzo for helpful discussions.
	We also thank Shiho Kobayashi for useful comments and the anonymous referee for a constructive report.

	\label{lastpage}
\end{document}